\documentclass{vldb}
\usepackage{latexsym}
\usepackage{multirow}
\usepackage{epsfig}
\usepackage{amsfonts}
\usepackage{amssymb}
\usepackage{amsmath}
\usepackage{times}
\usepackage[tight]{subfigure}
\usepackage{algorithm,algorithmic}
\usepackage{color}
\newcommand{\rw}[2]{{\color{black}#2}}
\newcommand{\x}[1]{{\color{red}\bf{}}}
\newcommand{\comment}[1]{}

\newtheorem{example}{Example}[section]

\newsavebox{\boxone}
\newsavebox{\boxtwo}
\newsavebox{\boxthree}

\newlength{\narrow}
\newlength{\cnarrow}
\newcommand{\topline}{
  \hrule
  \vskip .5\baselineskip}
\newcommand{\bottomline}{
  \vskip 2pt
  \hrule}

\newcommand{\chbox}[2]{
  \hbox to #1{\hss\vtop{#2}\hss}}

\newcommand{\nchbox}[1]{
  \chbox{\narrow}{#1}}

\newcommand{\cnchbox}[1]{
  \chbox{\cnarrow}{#1}}

\newcommand{\fcode}[1]{
  
  \chbox{\textwidth}{\tgrind\input{#1}}}

\newcommand{\ncode}[1]{
  
  \chbox{\narrow}{\tgrind\input{#1}}}

\def\nfig#1#2#3{
  \vtop{\nchbox{#1}
  \hbox to\narrow{\parbox{\narrow}{\caption{#2}\label{#3}}}}}

\newcommand{\cncode}[1]{
  \chbox{\cnarrow}{\tgrind\input{#1}}}

\def\codefiggen[#1]#2#3#4#5#6{
  \begin{figure}[#1]
  #5
  \fcode{#2}
  \center\parbox{.9\textwidth}{\caption{#3}\label{#4}}
  #6
  \end{figure}}

\def\codefig[#1]#2#3#4{
  \codefiggen[#1]{#2}{#3}{#4}{}{}}

\def\codefigline[#1]#2#3#4{
  \codefiggen[#1]{#2}{#3}{#4}{\topline}{\bottomline}}

\def\doublefiggen[#1]#2#3#4#5#6#7#8#9{
  \begin{figure}[#1]
  #8
  \hbox to \textwidth{
  \nfig{#2}{#3}{#4}
  \hfil
  \nfig{#5}{#6}{#7}}
  #9
  \end{figure}}

\def\doublefig[#1]#2#3#4#5#6#7{
  \doublefiggen[#1]{#2}{#3}{#4}{#5}{#6}{#7}{}{}}

\def\doublefigline[#1]#2#3#4#5#6#7{
  \doublefiggen[#1]{#2}{#3}{#4}{#5}{#6}{#7}{\topline}{\bottomline}}

\def\doublecodefig[#1]#2#3#4#5#6#7{
  \doublefig[#1]{\ncode{#2}}{#3}{#4}{\ncode{#5}}{#6}{#7}}

\def\doublecodefigline[#1]#2#3#4#5#6#7{
  \doublefigline[#1]{\ncode{#2}}{#3}{#4}{\ncode{#5}}{#6}{#7}}

\newcommand{\codepair}[4]{\vbox{
  \hbox{\ncode{#1} \hfil \ncode{#3}}
  \vskip .3\baselineskip plus .3\baselineskip
  \hbox{\hbox to\narrow{#2\hfil} \hfil \hbox to\narrow{#4\hfil}}}}

\def\codepairfig[#1]#2#3#4#5#6#7{
  \begin{figure}[#1]
  \codepair{#2}{#3}{#4}{#5}
  \center\parbox{.9\textwidth}{\caption{#6}}
  \label{#7}
  \end{figure}}

\def\cncodepairfiggen[#1]#2#3#4#5#6#7{
  \begin{figure}[#1]
  #6
  \hbox{\cncode{#2}\hfil\cncode{#3}}
  \center\parbox{.9\columnwidth}{\caption{#4}\label{#5}}
  #7
  \end{figure}}

\def\cncodepairfig[#1]#2#3#4#5{
  \cncodepairfiggen[#1]{#2}{#3}{#4}{#5}{}{}}

\def\cncodepairfigline[#1]#2#3#4#5{
  \cncodepairfiggen[#1]{#2}{#3}{#4}{#5}{\topline}{\bottomline}}

\def\doublefigOnecap*[#1]#2#3#4#5{
  \begin{figure*}[#1]
  \hbox to \textwidth{
  \nchbox{#2}
  \hfil
  \nchbox{#3}}
  \caption{#4}
  \label{#5}
  \end{figure*}}

\def\doublefigOnecap[#1]#2#3#4#5{
  \begin{figure}[#1]
  \topline
  \hbox to \columnwidth{
  \cnchbox{#2}
  \hfil
  \cnchbox{#3}}
  \caption{#4}
  \label{#5}
  \bottomline
  \end{figure}}

\def\PSfig[#1]#2#3#4{
 \begin{figure}
 \centerline{\psfig{file=#2,width=\columnwidth}}
 \caption{{#3}}
 \label{#4}
 \end{figure}}

\def\PSfiglines[#1]#2#3#4{
 \begin{figure}[#1]
 \topline
 \centerline{\psfig{file=#2,width=\columnwidth}}
 \caption{{#3}}
 \label{#4}
 \bottomline
 \end{figure}}

\def\PSfiglinesht[#1]#2#3#4#5{
 \begin{figure}[#1]
 \topline
 \centerline{\psfig{file=#2,height=#3}}
 \caption{{#4}}
 \label{#5}
 \bottomline
 \end{figure}}

\def\doublePSfig[#1]#2#3#4#5#6{
  \doublefigOnecap[#1]
    {\cnchbox{\psfig{file=#2,height=#4}}}
    {\cnchbox{\psfig{file=#3,height=#4}}}
    {#5}
    {#6}}

\newlength{\boxwidth}
\newcommand{\bproof}{{\bf Proof Sketch:}}
\newcommand{\eproof}{\mbox{$\Box$}}

\def\tabdoublecode#1#2#3#4{
 \begin{figure*}[t]
 \topline\vs{-.4}
 \hbox to \columnwidth{
 \vtop{\small
 \begin{tabbing}
 #1
 \end{tabbing}}
 \hfil
 \hfil
 \hfil
 \vtop{\small
 \begin{tabbing}
 #2
 \end{tabbing}}
 }
 \caption{#3\label{#4}}
 \bottomline
 \end{figure*}
}
\def\tabtriplecode#1#2#3#4#5{
 \begin{figure}
 \topline\vs{-.4}
 \hbox to \columnwidth{
 \vtop{\small
 \begin{tabbing}
 #1
 \end{tabbing}}
 \hfil
 \hfil
 \hfil
 \vtop{\small
 \begin{tabbing}
 #2
 \end{tabbing}}
 \hfil
 \hfil
 \hfil
 \vtop{\small
 \begin{tabbing}
 #3
 \end{tabbing}}
 }
 \caption{#4\label{#5}}
 \bottomline
 \end{figure}
}

\newtheorem{lemma}{Lemma}
\newcommand{\blemma}{\begin{lemma}}
\newcommand{\elemma}{\end{lemma}}

\newtheorem{thm}{Theorem}
\newcommand{\bthm}{\begin{thm}}
\newcommand{\ethm}{\end{thm}}

\newtheorem{defin}{Definition}
\newcommand{\bdefin}{\begin{defin}}
\newcommand{\edefin}{\end{defin}}

\newtheorem{observation}{Observation}
\newcommand{\bobserv}{\begin{observation}}
\newcommand{\eobserv}{\end{observation}}

\newcommand{\vs}[1]{\vspace{#1cm}}
\newcommand{\be}{\begin{equation}}
\newcommand{\ee}{\end{equation}}

\newcommand{\bdesc}{\begin{description}}
\newcommand{\edesc}{\end{description}}
\newcommand{\benum}{\begin{enumerate}}
\newcommand{\eenum}{\end{enumerate}}
\newcommand{\bitem}{\begin{itemize}}
\newcommand{\eitem}{\end{itemize}}
\newcommand{\bcenter}{\begin{center}}
\newcommand{\ecenter}{\end{center}}
\newcommand{\btabular}{\begin{tabular}}
\newcommand{\etabular}{\end{tabular}}
\newcommand{\beqnarr}{
 \begin{eqnarray}}
\newcommand{\eeqnarr}{\end{eqnarray}}

\title{Simple, Fast, and Scalable Reachability Oracle}

\numberofauthors{2}
\author{
\alignauthor
Ruoming Jin \\
      \affaddr{Department of Computer Science}\\
      \affaddr{Kent State University}\\
      \email{jin@cs.kent.edu}
\alignauthor
Guan Wang \\
      \affaddr{Department of Computer Science}\\
      \affaddr{Kent State University}\\
      \email{gwang@cs.kent.edu}
}

\date{}

\begin{document}
\maketitle

\begin{abstract}
A reachability oracle (or hop labeling) assigns each vertex $v$  two sets of vertices: $L_{out}(v)$ and $L_{in}(v)$, such that $u$ reaches $v$ iff $L_{out}(u) \cap L_{in}(v)\neq \emptyset$. 
Despite their simplicity and elegance, reachability oracles have failed to achieve efficiency in more than ten years since their introduction: the main problem is high construction cost, which stems from a set-cover framework and the need to materialize transitive closure. In this paper, we present two simple and efficient labeling algorithms, {\em Hierarchical-Labeling} and {\em Distribution-Labeling}, which can work on massive real-world graphs: their construction time is an order of magnitude faster than the set-cover based labeling approach, and transitive closure materialization is not needed. On large graphs, their index sizes and their query performance can now beat the state-of-the-art transitive closure compression and online search approaches. 

\comment{
Most of the existing reachability index performs well on small to medium size graphs. 
However, most of them have scalability bottleneck and scale to graphs with millions of vertices/edges.
As the size of graph is becoming increasingly large, scalability is quickly becoming the major research challenge
the reachability computation has to face today.  Can we construct
indices which can scale to very large graphs with tens of millions
of vertices and edges? Can the existing reachability indices
which performs well on the moderate size graphs be scaled to very
large graphs?  In this paper, we propose {\bf SCARAB} (stands for SCAlable ReachABility),
a unified reachability computation framework: it not only can scale
the existing state-of-art reachability indices, which otherwise can
only be constructed and work on moderate size graphs, but also
can significantly speedup the online query answering approaches. 
Our experimental results demonstrate the SCARAB approach can perform on 
graphs with tens of millions of vertices/edges and is also much faster then 
GRAIL, the state-of-art scalability index approach.
}
\end{abstract}

\vspace*{-1.0ex}
\section{Introduction}
\label{sec:intro}

As one of the most fundamental graph operators, reachability has drawn much research interest in recent years ~\cite{Chen05,Wang06,ChengYLWY06,Trissl07,DBLP:conf/sigmod/JinXRW08,DBLP:conf/edbt/ChengYLWY08,DBLP:conf/icde/ChenC08,DBLP:conf/sigmod/JinXRF09,Zhu:2009, yildirim:grail,Cai:2010,vanSchaik:2011,DBLP:journals/tods/JinRXW11,DBLP:conf/icde/ChenC11} and seems to continue fascinating researchers with new focuses~\cite{Jin:2012:SSR,Zhang:2012:ICM,Jin:2011:DRC} and new variants ~\cite{Fan:2012:PGD,Cheng:2012:KYS,Shirani-Mehr:2012:ERQ}.
The basic reachability query answers whether a vertex $u$ can reach another vertex $v$ using a simple path ($? u \rightarrow v$) in a directed graph.
It has a wide range of applications from software engineering, to distributed computing, to biomedical and social network analysis, to XML and the semantic web, among others.

The majority of the existing reachability computation approaches belong to either transitive closure materialization (compression) ~\cite{SIGMOD:AgrawalBJ:1989,Nuutila:1995,Wang06,DBLP:conf/sigmod/JinXRW08,vanSchaik:2011} or online search~\cite{Chen05,Trissl07,yildirim:grail}. The transitive closure compression approaches tend to be faster but generally have difficulty scaling to massive graphs due to the precomputation and/or memory cost. Online search is (often one or two orders of magnitude) slower but can work on large graphs~\cite{yildirim:grail,Jin:2012:SSR}.
The latest research ~\cite{Jin:2012:SSR} introduces a unified SCARAB method based on ``reachability backbone'' (similar to the highway in the transportation network) to deal with their limitations: it can both help scale the transitive closure approaches and speed up online search. However, the query performance of transitive closure approaches tends to be slowed down and they may still not work if the size of the reachability backbone remains too large~\cite{Jin:2012:SSR}.

The reachability oracle, more commonly known as hop labeling, ~\cite{cohen2hop,Thorup:2004:COR} is an interesting third category of approaches which lie between transitive closure materialization and online search.
Each vertex $v$ is labeled with two sets: $L_{out}(v)$, which contains hops (vertices) $v$ can reach; and $L_{in}(v)$, which contains hops that can reach $v$.
Given $L_{out}(u)$ and $L_{in}(v)$, but nothing else, we can compute if $u$ reaches $v$ by determining whether there is at least a common hop, $L_{out}(u) \cap L_{in}(v)\neq \emptyset$.
The idea is simple, elegant, and seems very promising: hop labeling can be considered as a factorization of the binary matrix of transitive closure; thus it should be able to deliver more compact indices than the transitive closure and also offer fast query performance.

Unfortunately, after more than ten years since its first proposal ~\cite{cohen2hop} and a list of worthy attempts~\cite{hopiedbt,ChengYLWY06,DBLP:conf/edbt/ChengYLWY08, DBLP:conf/sigmod/JinXRF09,Cai:2010}, hop labeling or reachability oracle, still eludes us and still fails to meet its expectations.
Despite its appealing theoretical nature, recent studies ~\cite{Jin:2012:SSR,vanSchaik:2011,Cheng:2012:KYS,yildirim:grail} all seem to confirm its inability to handle real-world large graphs: hop labeling is expensive to construct, taking much longer time than other approaches, and can barely work on large graphs, due to prohibitive memory cost of the construction algorithm. Many studies~\cite{Jin:2012:SSR,vanSchaik:2011,Cheng:2012:KYS,yildirim:grail} also show up to an order of magnitude slower query performance compared with the fastest transitive closure compression approaches (though we discover the underlying reason is mainly due to the implementation of  hop labeling $L_{out}$ and $L_{in}$; employing a sorted vector/array instead of a set can significantly eliminate the query performance gap).

The high construction cost of the reachability oracle is inherent to the existing labeling algorithms and directly results in the scalability bottleneck.
In order to minimize the labeling size, many algorithms~\cite{cohen2hop,hopiedbt,ChengYLWY06, DBLP:conf/sigmod/JinXRF09,Cai:2010} rely on a greedy set-cover procedure, which involves two costly operators:
1) repetitively finding densest subgraphs from a large number of bipartite graphs; and 2) materialization of the entire transitive closure. The latter is needed since each reachability pair needs to be explicitly covered by a selected hop.
Even with concise transitive closure representation, such as using geometric format~\cite{ChengYLWY06}, or reducing the covered pairs using $3$-hop~\cite{DBLP:conf/sigmod/JinXRF09,Cai:2010}, the overall construction complexity is still close to or more than $O(n^3)$, which is still too expensive for large graphs.
Alternative labeling algorithms ~\cite{Thorup:2004:COR,DBLP:conf/edbt/ChengYLWY08} try to use graph separators, but only special graph classes, such as planar graphs, consisting of small graph separators, can adopt such techniques well~\cite{Thorup:2004:COR}.
For general graphs, the scalability of such approach~\cite{DBLP:conf/edbt/ChengYLWY08} is limited by the lack of good scalable partition algorithms for discovering graph separators on large graphs.

Can the reachability oracle be practical? Is it a purely theoretical concept which can only work on small toy graphs, or it is a powerful tool which can shape reality and can work on real-world large graphs with millions of vertices and edges?  Arguably, this is one of the most important unsolved puzzles in reachability computation.
This work resolves these questions by presenting two simple and efficient labeling algorithms, {\em Hierarchical-Labeling} and {\em Distribution-Labeling}, which can work on massive real-world graphs. Their construction costs are as fast as the state-of-the-art transitive closure compression approaches, there is no expensive transitive closure materialization, dense subgraph detection, or greedy set-cover procedure, there is no need for graph separators, and on large graphs, their index sizes and their query performance beat the state-of-the-art transitive closure compression and online search approaches~\cite{DBLP:conf/sigmod/JinXRW08,vanSchaik:2011,Jin:2012:SSR,vanSchaik:2011,Cheng:2012:KYS,yildirim:grail}.
Using these two algorithms, the power of hop labeling is finally unleashed and a fast,  compact and scalable reachability oracle becomes a reality.
%These two algorithms unleash the powerful of hop labeling and are the first to realize a fast, compact and scalable reachability reachability oracle based on our best knowledge.

\comment{
The rest of the paper is organized as follows.
In Section~\ref{related}, we review the prior work on reachability.
In Section~\ref{basic}, we give an overview of the basic ideas of constructing a fast, compact and scalable reachability oracle.
In Section~\ref{hierarchy}, we present the {\em Hierarchical-Labeling} algorithm, which is based on a hierarchical decomposition of a DAG (direct acyclic graph).
In Section~\ref{order}, we introduce the {\em Distribution-Labeling} algorithm, which utilizes a total vertex order.
In Section~\ref{expr}, we report the detailed experimental study on these two new labeling algorithms compared with the state-of-the-art reachability computation approaches.
We offer concluding remarks in Section~\ref{conc}.
}

\vspace*{-2.0ex}
\section{Related Work}
\label{related}

To compute the reachability, the directed graph is typically transformed  into a DAG (directed acyclic graph)  by coalescing strongly connected components into vertices, avoiding the trivial case where vertices reach each other in a strongly connected component.
The size of the DAG is often much smaller than that of the original graph and is more convenient for reachability indexing.
Let $G=(V,E)$ be the DAG for a reachability query, with number of vertices $n=|V|$ and number of edges $m=|E|$.

\vspace*{-1.0ex}
\subsection{Transitive Closure and Online Search}
There are two extremes in computing reachability. At one end, the entire transitive closure ($TC$) of $G$ is precomputed and fully materialized (often in a binary matrix). Since the reachability between any pair is recorded,  reachability can be answered in constant time, though the $O(n^2)$ storage is prohibitive for large graphs.
At the other end, DFS/BFS can be employed. Though it does not need an additional index, its query answering time is too slow for large graphs.
As we mentioned before, the majority of the reachability computation approaches aim to either compress the transitive closure~\cite{SIGMOD:AgrawalBJ:1989,Jagadish90,Nuutila:1995,Wang06,DBLP:conf/sigmod/JinXRW08,DBLP:conf/icde/ChenC11,vanSchaik:2011,DBLP:journals/tods/JinRXW11} or to speed up the online search~\cite{Chen05,Trissl07,yildirim:grail}.

\noindent{\bf Transitive Closure Compression:}
This family of approaches aims to compress the transitive closure --
each vertex $u$ records a compact representation of $TC(u)$, i.e.,  all the vertices it reaches.
The reachability from vertex $u$ to $v$ is computed by checking vertex $v$ against $TC(u)$.
Representative approaches include chain compression~\cite{Jagadish90,DBLP:conf/icde/ChenC08}, interval or tree compression~\cite{SIGMOD:AgrawalBJ:1989,Nuutila:1995}, dual-labeling~\cite{Wang06},  path-tree~\cite{DBLP:conf/sigmod/JinXRW08}, and bit-vector compression~\cite{vanSchaik:2011}.
Using interval-compress as an example,  any contiguous vertex segment in the original $TC(u)$ is represented by an interval.
For instance, if $TC(u)$ is $\{1,2,3,4,8,9,10\}$, it can be represented as two intervals: $[1,4]$ and $[8,10]$.

Existing studies ~\cite{vanSchaik:2011,yildirim:grail,Jin:2012:SSR} have shown these approaches are the fastest in terms of query answering since checking against transitive closure $TC(u)$ is typically quite simple (linear scan or binary search suffices);  in particular, the interval and path-tree approaches seem to be the best in terms of query answering performance. However, the transitive closure materialization, despite compression, is still costly.
The index size is often the reason these approaches are not scalable on large graphs~\cite{yildirim:grail,Jin:2012:SSR}.

\noindent{\bf Fast Online Search:}
Instead of materializing the transitive closure, this set of approaches~\cite{Chen05,Trissl07,yildirim:grail} aims to speed up the online search.
To achieve this,  auxiliary labeling information per vertex is precomputed and utilized for pruning the search space.
Using the state-of-the-art GRAIL ~\cite{yildirim:grail} as an example, each vertex is assigned multiple interval labels where each interval is computed by a random depth-first traversal.
The interval can help determine whether a vertex in the search space can be immediately pruned because it never reaches the destination vertex $v$.

The pre-computation of the auxiliary labeling information in these approaches is generally quite light; the index size is also small. Thus, these approaches can be applicable to very large graphs.
However, the query performance is not appealing; even the state-of-the-art GRAIL can be easily one or two orders of magnitude slower than the fast interval and path-tree approaches~\cite{yildirim:grail,Jin:2012:SSR}.
For very large graphs, these approaches may be too slow for answering reachability query.

\vspace*{-1.0ex}
\subsection{Reachability Oracle}

The reachability oracle~\cite{cohen2hop,Thorup:2004:COR}, also refer to as hop labeling, was pioneered by Cohen {\em et al.}~\cite{cohen2hop}.
Though it also encodes transitive closure, it does not explicitly compress the transitive closure of each individual vertex independently (unlike the transitive closure compression approaches).
Here,   each vertex $v$ is labeled with two sets: $L_{out}(v)$, which contains hops (vertices) $v$ can reach; and $L_{in}(v)$, which contain hops that can reach $v$.
Given $L_{out}(u)$ and $L_{in}(v)$, but nothing else, we can compute if $u$ reaches $v$ by determining whether there is a common hop, $L_{out}(u) \cap L_{in}(v)$.
In fact, a reachability oracle can be considered as a factorization of the binary matrix of transitive closure~\cite{DBLP:conf/sigmod/JinXRF09}; and thus more compact indices are expected from such a scheme.

The seminal $2$-hop labeling~\cite{cohen2hop} aims to minimize the reachability oracle size, which is the total label size $\sum (|L_{out}(u)|+|L_{in}(u)|)$.
It employs an approximate (greedy) algorithm based on set-covering which
can produce a reachability oracle with size no larger than the optimal one by a logarithmic factor.
The optimal $2$-hop index size is conjectured to be $\tilde{O}(nm^{1/2})$.
The major problem of the $2$-hop indexing approach is its high construction cost, which needs to iteratively find dense subgraphs from a large number of bipartite graphs (representing the covering of transitive closure). 
Its computational cost is $O(n^3 |TC|)$, where $|TC|$ is the total size of transitive closure.
A number of approaches have sought to reduce construction cost through speeding up the set cover procedure~\cite{hopiedbt}, using concise transitive closure representation~\cite{ChengYLWY06}, or reducing the covered pairs using $3$-hop~\cite{DBLP:conf/sigmod/JinXRF09,Cai:2010}. 
However, they still need to repetitively find densest subgraphs and to materialize the transitive closure. 
Alternative labeling algorithms ~\cite{Thorup:2004:COR,DBLP:conf/edbt/ChengYLWY08} try to use graph separators, but only special graph classes, such as planar graphs, consisting of small graph separators, can adopt such technique well~\cite{Thorup:2004:COR}.
For general graphs, the scalability of such approach~\cite{DBLP:conf/edbt/ChengYLWY08} is limited by the lack of good scalable partition algorithms for discovering graph separators on large graphs.

\comment{
{\em The greedy set-covering algorithm needs to iteratively find a vertex $v$ associated with two subsets of vertices $X$ and $Y$ which utilizes $v$ as the intermediate hop, i.e., $v \in L_{out}(x), x \in X$ and $v \in L_{in}(y), y \in Y$.
To select vertex $v$ and its associated $X$ and $Y$, the greedy procedure utilizes {\bf price}, which measures the cost-benefit tradeoff between recording the vertex in $L_{out}(x)$ and $L_{in}(y)$ (cost) and the number of reachability pairs being newly covered (benefit) by such labeling: $\frac{|X|+|Y|}{|X \times Y \setminus C|}$, where $C$ are those reachable pairs already covered by previously selected hops.}
This selection step can be transformed into the problem of finding a densest subgraph in $n$ bipartite graphs.
The approximate algorithm to solve this subproblem is in the linear order with respect to the number of edges in the bipartite graph.
Such an iterative approach can be as costly as $O(n^3 |TC|)$, where $|TC|$ is the total size of transitive closure.

A number of approaches have sought to reduce construction cost through speeding up the set cover procedure~\cite{hopiedbt}, using concise transitive closure representation~\cite{ChengYLWY06}, or reducing the covered pairs using $3$-hop~\cite{DBLP:conf/sigmod/JinXRF09,Cai:2010}.
However, they still need to repetitively find densest subgraphs from a large number of bipartite graphs and to materialize the transitive closure to explicitly confirm each reachable pair is covered by the hop labeling.
Alternative labeling algorithms ~\cite{Thorup:2004:COR,DBLP:conf/edbt/ChengYLWY08} try to use graph separators, but only special graph classes, such as planar graphs, consisting of small graph separators, can adopt such technique well~\cite{Thorup:2004:COR}.
For general graphs, the scalability of such approach~\cite{DBLP:conf/edbt/ChengYLWY08} is limited by the lack of good scalable partition algorithms for discovering graph separators on large graphs.}

\vspace*{-1.0ex}
\subsection{Reachability Backbone and SCARAB}
\label{substructure}
%Interestingly, two latest studies ~\cite{Jin:2012:SSR,Cheng:2012:KYS} both consider to utilize certain substructures (graph minors, where a graph minor $H$ can be obtained from graph $G$ by edge deletion, edge contraction and vertex %removal~\cite{Diestel00}) to help with reachability computation though their focuses and techniques are quite different.
%Note that graph minor will preserve some of the reachability information in the original graph but importantly, it will not add any new reachability.

In the latest study~\cite{Jin:2012:SSR}, the authors introduce a general framework, referred to as SCARAB (SCAling ReachABility), for scaling the existing reachability indices (including both transitive closure compression and hop labeling approaches) and for speeding up the online search approaches. 
The central idea is to leverage a ``reachability  backbone'', which carries the major ``reachability flow'' information. The reachability backbone \rw{shares the similar spirit of}{is similar in spirit to} the highway structure used in several state-of-the-art shortest path distance computation methods on road networks~\cite{BastFSS07,Sanders05,Sanders:2012:EHH}. 
However, the SCARAB work~\cite{Jin:2012:SSR} is one of the first studies to construct and utilize such structure in the reachability computation. 

%In other words, reachability backbone serves as the highway of the original graph and the existing reachability indices can be performed on the reachability backbone instead of the original graph.

Formally, the reachability backbone $G^\star=(V^\star, E^\star)$ of graph $G$ is defined as a subgraph of the transitive closure of $G$  ($E^\star \subseteq TC(G)$),  such that {\em for any reachable $(u,v)$ pair, there must exist local neighbors $u^\star \in V^\star$, $v^\star \in V^\star$ with respect to locality threshold $\epsilon$, i.e., $d(u,u^\star)\leq \epsilon$ and $d(v^\star,v) \leq \epsilon$, and $u^\star \rightarrow v^\star$.} Here $d(u,u^\star)$ is the shortest path distance from $u$ to $u^\star$ where the weight of each edge is unit. 
To compute the reachability from $u$ to $v$, $u$ collects a list of local outgoing backbone vertices (entries) using forward BFS, and $v$ collects a list of local incoming backbone vertices (exits) using backward BFS. Then an existing reachability approach can be utilized to determine if there is a local entry reaching a local exit on the reachability backbone $G^\star$.

Two algorithms are developed to approximate the minimal backbone, one based on set-cover and the other based on BFS. The latter, referred to as {\em FastCover}, is particularly efficient and effective, with time complexity $O(\sum_{v \in V} |N_\epsilon(v)| log |N_\epsilon(v)| + |E_\epsilon(v)|)$, where $N_\epsilon(v)$ ($E_\epsilon(v)$) is the set of vertices (edges) $v$ can reach in $\epsilon$ steps. Experiments show that even with $\epsilon$, the size of the reachability backbone is significantly smaller than the original graph (about $1/10$ the number of vertices of the original graph). As we will discuss later, our first {\em Hierarchical-Labeling} algorithm is directly inspired by the reachability backbone and effectively utilizes it for reachability oracle construction. 

Though the scaling approach is quite effective for helping deal with large graphs, it is still constrained by the power of the original index approaches. For many large graphs, the reachability backbone can still be too large for them to process as shown in the experiment study in ~\cite{Jin:2012:SSR}. Also, using the reachability backbone slows down the query performance of the transitive closure compression and hop labeling approaches (typically two or three times slower than the original approaches) on the graphs where they can still run. In addition, theoretically, the reachability backbone could be applied recursively; this may further slow down query performance. In ~\cite{Jin:2012:SSR}, this option is not studied.

We also note that in ~\cite{Cheng:2012:KYS}, a new variant of reachability queries, $k$-hop reachability, is introduced and studied. It asks whether vertex $u$ can reach $v$ within $k$ steps. This problem can be considered a generalization of the basic reachability, where $k= \infty$. A $k$-reach indexing approach is developed and the study shows that approach can handle basic reachability quite effectively (with comparable query performance to the fastest transitive closure compression approaches on small graphs).
The $k$-reach indexing approach is based on vertex cover (a set of vertices covers all the edges in the graph), and it actually produces  a reachability backbone with $\epsilon=1$ as defined in ~\cite{Jin:2012:SSR}. 
But this study directly materializes the transitive closure between any pair of vertices in the vertex cover, where in ~\cite{Jin:2012:SSR}, the existing reachability indices are used.
Thus, for very large graphs where the vertex cover is often large, the pair-wise reachability materialization is not feasible. 

%(This observation is also confirmed through our experimental study in Section~\ref{expr}).
%To compute the reachability from $u$ to $v$, each vertex only needs to access their immediate neighbors in the vertex cover; and the pairwise reachability between any two vertices in the set cover is precomputed and fully materialized (for basic reachability computation). It is easy to see that {\em this vertex cover based approach is a reachability backbone with $\epsilon=1$ as defined in ~\cite{Jin:2012:SSR}}.

\vspace*{-1.0ex}
\subsection{Other Related Works}

\noindent{\bf Distance $2$-HOP Labeling:}
The 2-hop labeling method proposed by Cohen {\em et al.}~\cite{cohen2hop} can also handle the exact distance labeling. 
Here, each vertex $u$ records a list of intermediate vertices $OUT(u)$ which it can reach along with their (shortest) distances, and a list of intermediate vertices $IN(u)$ which can reach it along with their distances.
To answer the point-to-point shortest distance query from $u$ to $v$,
we simply need to check all the common intermediate vertices between $OUT(u)$ and $IN(v)$ and choose the vertex $p$, such that $dist(u,p)+dist(p,v)$ is minimized for all $p \in OUT(u) \cap IN(v)$.
However, its computational cost (similar to the reachability $2$-hop labeling) is too expensive even for graphs with hundreds of thousands of vertices.  

Recently, Abraham {\em et al.}~\cite{Abraham:2011:HLA} have developed a fast and practical algorithm to heuristically construct the distance labeling on large road networks. 
In particular, they utilize {\em contraction hierarchies} (CH)~\cite{Geisberger:2008:CHF} which transform the original graph into a level-wise structure, and then assign the maximum-rank vertex on the shortest path between $s$ and $t$ as the hop for $s$ and $t$. 
However, the core of CH needs to iteratively remove vertices and then add {\em shortcuts} for fast shortest path computation. Due to the power-law property, such operation easily becomes very expensive for general graphs.
% (considering a vertex with a thousand of neighbors, and removing it may need to check millions of potential shortcuts).
For example, to remove a vertex with thousands of neighbors may require checking millions of potential shortcuts.
Interestingly, another state-of-the-art method, {\em Path-Oracle}~\cite{Sankaranarayanan:2009} by Sankaranarayanan {\em et al.}, utilizes a spatial data structure for distance labeling on road networks.   
In ~\cite{DBLP:conf/sigmod/JinRXL12}, we proposed a highway-centric labeling approach to label large sparse graphs. The basic idea is to utilize a highway structure, such as a spanning tree, to reduce the computational cost of labeling as well as to reduce the labeling size. However, it still has a scalability bottleneck as it needs to partially materialize the transitive closure for directed graphs.

\noindent{\bf Relationship to the latest reachability labeling ~\cite{DBLP:conf/sigmod/ChengHWF13} and distance labeling ~\cite{DBLP:conf/sigmod/AkibaIY13} papers:}

We have recently become aware that Cheng {\em et al.}~\cite{DBLP:conf/sigmod/ChengHWF13} have developed a reachability labeling approach, referred to as {\em TF-label}. Their approach is similar to the {\em Hierarchical Labeling} (HL) approach being introduced in this work.  In particular, it can be considered a special case of HL where $\epsilon=1$ (Section ~\ref{hierarchy}).  The hierarchy being constructed in ~\cite{DBLP:conf/sigmod/ChengHWF13} is based on iteratively extracting a reachability backbone with $\epsilon=1$, inspired by independent sets.  A similar approach has been used in their earlier work on distance labeling, referred to as IS-labeling ~\cite{DBLP:journals/corr/abs-1211-2367}. In this paper, the hierarchy structure is extracted based on the reachability backbone approach~\cite{Jin:2012:SSR}, which has been shown to be effective and efficient for scaling reachability computation. 
In another recent work~\cite{DBLP:conf/sigmod/AkibaIY13}, Akiba {\em et al.} have proposed a distance labeling approach, referred to as the {\em Pruned Landmark}. This approach is similar in spirit to the {\em Distribution Labeling} (DL) approach. However, DL performs BFS in both directions (forward and reverse) in order to handle reachability labeling. Also, the condition for assigning labels is different. 

Finally, we would like to point out that both Hierarchical Labeling (HL) and Distribution Labeling (DL) are proposed independently of  ~\cite{DBLP:conf/sigmod/ChengHWF13} and ~\cite{DBLP:conf/sigmod/AkibaIY13}. Indeed, even with both papers being accepted earlier than this work, this work is still novel: 1) the Hierarchical Labeling (HL) algorithm is a  general framework which captures ~\cite{DBLP:conf/sigmod/ChengHWF13} as a special instance; 2) the Distribution Labeling (DL) algorithm is not mentioned in ~\cite{DBLP:conf/sigmod/ChengHWF13} and is inspired from HL; 3) the Distribution Labeling (DL) algorithm, though similar in spirt to  ~\cite{DBLP:conf/sigmod/AkibaIY13}, is the first to deal with reachability. Also, the proof of labeling completeness  is non-trivial and its non-redundancy property is not studied in ~\cite{DBLP:conf/sigmod/AkibaIY13} (Section~\ref{order}).

\vspace{-2.0ex}
%\newpage
\section{Approach Overview}
\label{basic}

In a reachability oracle of graph $G$, each vertex $v$ is labeled with two sets: $L_{out}(v)$, which contains hops (vertices) $v$ can reach; and $L_{in}(v)$, which contain hops that can reach $v$.
A labeling is {\em complete} if and only if for any vertex pair where $u \rightarrow v$, $L_{out}(u) \cap L_{in}(v) \neq \emptyset$.
%i.e., there is at least a common hop such that $u$ reaches (recorded in $L_{out}(u)$) and it reaches $v$ (recorded in $L_{in}(v)$. 
The goal is to minimize the total label size, i.e., $\sum (|L_{out}(u)|+|L_{in}(u)|)$.
A smaller reachability oracle not only help to fit the index in main memory, but also speeds up the query processing (with $O(|L_{out}(u)|+|L_{in}(v)|)$ time complexity). 

As we mentioned before, though the existing set-cover based approaches~\cite{cohen2hop,hopiedbt,ChengYLWY06,DBLP:conf/sigmod/JinXRF09,Cai:2010} 
can achieve approximate optimal labeling size within a logarithmic factor, its computational and memory cost is prohibitively expensive for large graphs. The labeling process not only needs to materialize the transitive closure, but it also uses an iterative set-cover procedure which repetitively invokes dense subgraph detection. 
The reason for such complicated algorithm is that the following two criteria need to be met: 
1) a labeling must be complete, and
%has to be complete, i.e., any reachable pair do record a common hop; 
2) we wish the labeling to be minimal.
%, i.e., the right hop needs to be recorded by the corresponding vertices. 
The existing approach~\cite{cohen2hop,DBLP:conf/sigmod/JinXRF09} essentially transforms the labeling problem into a set cover problem with the cost of constructing the ground set (which is the entire transitive closure) and dynamic generation and selection of good candidate sets (through dense subgraph detection).
 
To achieve efficient labeling which can work on massive graphs, the following issues have to appropriately handled: 

\noindent{\bf 1. (Completeness without Transitive Closure):} 
Can we guarantee labeling completeness without materialization of the transitive closure?
Even compact ~\cite{ChengYLWY06} or reduced ~\cite{DBLP:conf/sigmod/JinXRF09} materialization can be expensive for large graphs. Thus, the key is whether a labeling process can avoid the need to explicitly check whether a reachable pair (against some form of transitive closure) is covered by the existing labeling. 
%Interestingly, we note that the complete transitive closure itself can be considered a complete distance oracle, i.e., $L_{out}(u)=TC(u)$ and $L_{in}(u)=\{u\}$. However, its label size is not very meaningful. This leads to the next problem. 

\noindent{\bf 2. (Compactness without Optimization):} 
Without the set-cover, it seems difficult to produce bounded approximate optimal labeling. 
But this does not mean that a compact reachability oracle cannot be produced. 
Clearly, each vertex should not record every valid hop in the labeling.
In the set-cover framework,  a price is computed to determine whether a vertex should be added to certain vertex labels. 
What other criteria can help determine the importance of hops (vertices) so that each vertex can be more selective in what it records? 

\comment{
The problem is what properties can be used to help reduce the ``redundant'' and ``unbalanced'' labeling. 
If a hop $x$ can be removed from $L_{in}(v)$ or $L_{out}(v)$ and all the reachability related to $v$ is still preserved, then $x$ is a redundant. 
The aforementioned transitive closure based labeling is non-redundant, but it is very ``unbalanced'' as the hop is only recorded in one side ($L_{out}(u)$).  
Thus, to produce compact distance oracle, the labeling needs to be non-redundant and balanced. }

In this paper, we investigate how the hierarchical structure of a DAG can help produce a complete and compact reachability oracle. The  basic idea is as follows: assuming a DAG can be represented in a hierarchical (multi-level) structure, such that the lower-level reachability needs to go through upper-level (but vice versa), then we can somehow recursively broadcast the upper-level labels to lower-level labels. In other words, the labels of lower-level vertices ($L_{in}$ and $L_{out})$ can directly utilize the already computed labels in the upper-level. Thus, on one side, by using the hierarchical structure, the completeness of labeling can be automatically guaranteed.   On the other side,  
it provides an importance score (the level) of every hop; and each vertex only records those hops whose levels are higher than or equal to its own level. We note that there have been several studies~\cite{Shekhar97,Sanders05,Delling06,Kriegel:2008:HGE,Bauer:2010:CHG,Abraham:2011:HLA} using the hierarchical structure for shortest path distance computation on road networks; however, how to construct and utilize the hierarchical structure for reachability computation has not been fully addressed. To the best of our knowledge, this is the first study to construct a fast and scalable reachability oracle based on hierarchical DAG decomposition. 

Now, to turn such an idea into a fast labeling algorithm for reachability oracle, the following two research questions need to be answered: 1) What hierarchical structure representation of a DAG can be used? 2) How should $L_{out}$ and $L_{in}$ be computed efficiently using a given hierarchical structure? 
In this paper, we introduce two fast labeling algorithms based on different hierarchical structures of a DAG: 

\noindent{\bf Hierarchical-Labeling (Section~\ref{hierarchy}):} In this approach, the hierarchical structure is produced by  a recursive reachability backbone approach, i.e., finding a reachability backbone $G^\star$ from the original graph $G$ and then applying the backbone extraction algorithm on $G^\star$. Recall that the reachability backbone is introduced by the latest SCARAB framework~\cite{Jin:2012:SSR} which aims to scale the existing reachability computation approaches. Here we apply it recursively to provide a hierarchical DAG decomposition. Given this, a fast labeling algorithm is designed to quickly compute $L_{in}$ and $L_{out}$ one vertex by one vertex in a level-wise fashion (from higher level to lower level). 
 
\noindent{\bf Distribution-Labeling (Section~\ref{order}):} In this approach, the sophisticated reachability backbone hierarchy is replaced with the simplest hierarchy  -- a total order, i.e.., each vertex is assigned a unique level in the hierarchy structure. Given this, instead of computing $L_{in}$ and $L_{out}$ one vertex at a time, the labeling algorithm will {\em distribute} the hop one by one (from higher order to lower order) to $L_{in}$ and $L_{out}$ of other vertices. The worst case computation complexity of this labeling algorithm is $O(n(n+m))$ (of the same order as transitive closure computation), though in practice it is much faster than the transitive closure computation. 

In the experimental study (Section~\ref{expr}), through an extensive study on both real and synthetic graphs, we found that both labeling approaches  not only are fast (up to an order of magnitude faster than the best set-cover based approach~\cite{cohen2hop,DBLP:conf/sigmod/JinXRF09}) and work on massive graphs, but most surprisingly, their label sizes are actually smaller than the set-cover based approaches.

\comment{

As an extreme case, $L_{in}(v)$ can record all the vertices $v$ reaches, and $L_{out}(v)$ records 
For instance, if $L_{in}(v)$ records

But how a labeling process can be programmed in a way such that redundant labeling can be maximally avoided?  To define this, we would prefer a labeling such that no element can be removed; then it is a concise labeling. 
For instance, if we can consider a labeling is redundant if both of them can remove a vertices without affecting the completeness, then it is not concise. The labeling produced by set-cover framework is very close to a concise labeling as each label will cover some new elements. But it does not gurantee that the later labeling can actually eliminate an early labeling. Note that checking conciseness is more expensive than checking the completeness. Here we do not want to explicitly validate that, but it serves as a general criterion to produce good labeling. 
Can we generate a concise labeling?

In SCARAB, we demonstrated the Backbone structure can effectively reduce the label size and query time. However compute the reachability information among all Backbone vertices can still be expensive spatialy and temporally, therefore we proposed an approach called HierarchyTwohop to reduce that cost.

The basic idea of HierarchyTwohop is rather simple:
%\benum
%\vspace*{-1.0ex}

\noindent{\bf 1. (Build Hierarchy)} For any given graph, HierarchyTwohop uses the Backbone discovery algorithm proposed in SCARAB to build a Backbone structure, then use the generated Backbone as also given graph, and build another Backbone based on it. HierarchyTwohop repeats this procesure until reaches the limit that user defines and finally generated a Hierarchy with mutiple levels. The top level we call it a "core" graph, which nomally has hundreds of nodes according to our experiment result.

\noindent{\bf 2. (Label "core" graph)} Because of the small scale of the "core" graph, we choose Twohop algorithm to assign labels on this level.

\noindent{\bf 3. (Label lower levels)} We proposed a novel algorithm to label lower level graphs, that each vertex only needs to be computed once on the highest level that contains it.

The difference between Hierarchy and SCARAB is not only that there's multiple levels of Backbone structures in Hierarchy. Also in Hierarchy, the Backbones will only be used in labeling phase. We answer the reachability queries with just labels.

In design of Hierarchy, we need to consider one basic research problems:
\benum
\vspace*{-1.0ex}
\item {\em How can we design the labeling algorithm on lower levels, so that can minimize the spatial and temporally computation cost.}
\vspace*{-1.0ex}
\eenum

To answer this question, we prove a lemma that it's sufficient for each vertex to be just computed once on the highest level of Backbone that contains it.

Even though the performance of HierarchyTwohop is promising, it's still struggling with memory usage. After all we have to build and store the hierarchy. And during the process of developing HierarchyTwohop, we realize that it's advantage is based on a computational order of all vertices. which drives u to think if we can find an easier way to generate the order instead of building the hierarchy. Finally we came up with a simple yet effective algorithm called ScalableOracle.

}

\comment{
The remainder of the paper is organized as follows. Section~\ref{hierarchy} introduce the discovery and labeling of HierarchyTwohop.  Section~\ref{scalableoracle} focuses on the optimization algorithm ScalableOracle. Section~\ref{computation} discusses the usage reachability indices to quickly answer reachability join test. Section~\ref{expr} provides a detailed empirical study to demonstrate the efficiency and scalability of SCARAB.

Finally, the paper is concluded in Section~\ref{conc}.
% and even on small to moderate graphs (graphs with tens of thousands vertices and edges), SCARAB is quite comparable
}

\comment{
2) {\em How can we utilize it for efficiently answering reachability computation? }

}

\comment{
In this paper, we make the following contributions.

}

\comment{
In particularly, we analyze the tradeoff between query efficiency, the index size, and the construction cost of different methods.
Such analysis will shed the lights on their scalability bottleneck and how we may solve them.
However,

On one side, we can perform online search such as DFS/BFS, which does not need additional index and thus do not have scalability issue in terms of construction;

\subsection{Prior Work on Reachability Indexing}
\label{prior}
}

\comment{
  proposed to reducing the index size.
{\em Unfortunately, in order to minimize the index size under the respective scheme,
they nonetheless all require the information of the transitive closure ($TC$).
Thus, even when the final reachability index size is much smaller than $TC$ and
possibly can handle large graphs, the construction algorithm simply cannot. }
This constitutes to the very reason for most of the scalability problems in reachability indexing.
Some methods try to handle this problem through heuristic methods, however, their index size
cannot guarantee the optimality~\cite{}.

In Table~\ref{summarization}, we summarize  the state-of-the-art approaches in terms of their
{\em query time, index size, construction time, and whether the transitive closure $TC$ is required in the construction.}
Parameter $k$ is the width of the chain decomposition of DAG $G$~\cite{Jagadish90}, $t$ is the number of (non-tree) edges left after removing all the edges of a spanning tree of $G$~\cite{Wang06}, and $k^\prime$ is the width of the path decomposition~\cite{DBLP:conf/sigmod/JinXRW08}. These three parameters $k$, $t$ and $k^\prime$, are method-specific and will be explained in more detail when we discuss their corresponding methods.

\begin{table}
\label{summarization}
{\scriptsize
 \begin{tabular}{lllll}\hline
& Query Time & Index Size & Constr. Time & $TC$ \\ \hline\hline
Opt. Tree Cover~\cite{SIGMOD:AgrawalBJ:1989}& $O(\log{n})$ & $O(n^2)$ & $O(nm)$ & Yes \\
Dual Labeling~\cite{Wang06} & $O(1)$ &$O(n+l^2)$ & $O(n+m+l^3)$ & $O(l^2)$ \\
Path-Tree ~\cite{DBLP:conf/sigmod/JinXRW08} & $\log^2{k^\prime}$ & $O(nk^\prime)$ & \textcolor{black}{$O(mn)$} & Yes  \\ \hline
2-Hop ~\cite{cohen2hop} & $\tilde{O}(m^{1/2})$ & $\tilde{O}(nm^{1/2})$ & \textcolor{black}{${O}(n^3|TC|)$} & Yes \\
3-Hop  ~\cite{} & $\tilde{O}(m^{1/2})$ & $\tilde{O}(nm^{1/2})$  & \textcolor{black}{${O}(n^3|TC|)$} & O(|Con|) \\                    \hline
Labeling+SSPI~\cite{Chen05}&  $O(m-n)$ & $O(n+m)$ & $O(n+m)$ &  No  \\
GRIPP~\cite{Trissl07} & $O(m-n)$ & $O(m+n) $& $O(n+m)$&No \\
GRAIL~\cite{}  & $O(n+m)$ & $O(dn)$ & $O(d(n+m))$  & No \\ \hline
Mtree (this paper) & $O(1)$ & $O(t(k+2)n)$ & $O(t (m n^\prime))$ & No \\ \hline
\end{tabular}
}
%\vspace*{-3.0ex}
\caption{Worst-Case Complexity}
  \label{tab:cmp}
%\vspace*{-5.0ex}
\end{table}
}

\vspace*{-2.0ex}
\section{Hierarchical Labeling}
\label{hierarchy}

Before we proceed to discuss the {\em Hierarchical Labeling} approach, let us formally introduce the one-side reachability backbone (first defined in ~\cite{Jin:2012:SSR} for scaling the existing reachability computation), which serves as the basis for hierarchical DAG decomposition  and the labeling algorithm.

\bdefin ({\bf One-Side Reachability Backbone}~\cite{Jin:2012:SSR})
Given DAG $G$, and local threshold $\epsilon$, the one-side reachability backbone $G^\star=(V^\star, E^\star)$ is defined as follows: 1) $V^\star \subseteq V$, such that for any vertex pair $(u,v)$ in $G$ with $d(u,v)=\epsilon$, there is a vertex $v^\star$ with $d(u,v^\star) \leq \epsilon$ and $d(v^\star,v) \leq \epsilon$;
2) $E^\star$ includes the edges which link vertex pair $(u^\star,v^\star)$ in $V^\star$ with $d(u^\star,v^\star) \leq \epsilon+1$.
\edefin

Note that $E^\star$ can be simplified as a transitive reduction ~\cite{Jin:2012:SSR} (the minimal edge set preserving the reachability). Since computing transitive reduction is as expensive as transitive closure, rules like the following can be applied: $(u^\star,v^\star) \in E^\star$ can be removed if there is another intermediate vertex $x \in V^\star$ (not $u^\star$ and $v^\star$) with $d(u^\star,x) \leq \epsilon$ and $d(x,v^\star) \leq \epsilon$. To facilitate our discussion, for any two vertices $u$ and $v$, if their distance is no higher than $\epsilon$ (local threshold), we refer to them as being a local pair (or being local to one another). 

\begin{figure*}[t!]
    \label{fig:hierarchyexample}
    \centering
    \mbox{
        \subfigure[Original Graph $G_0$]{\includegraphics[scale=0.2]{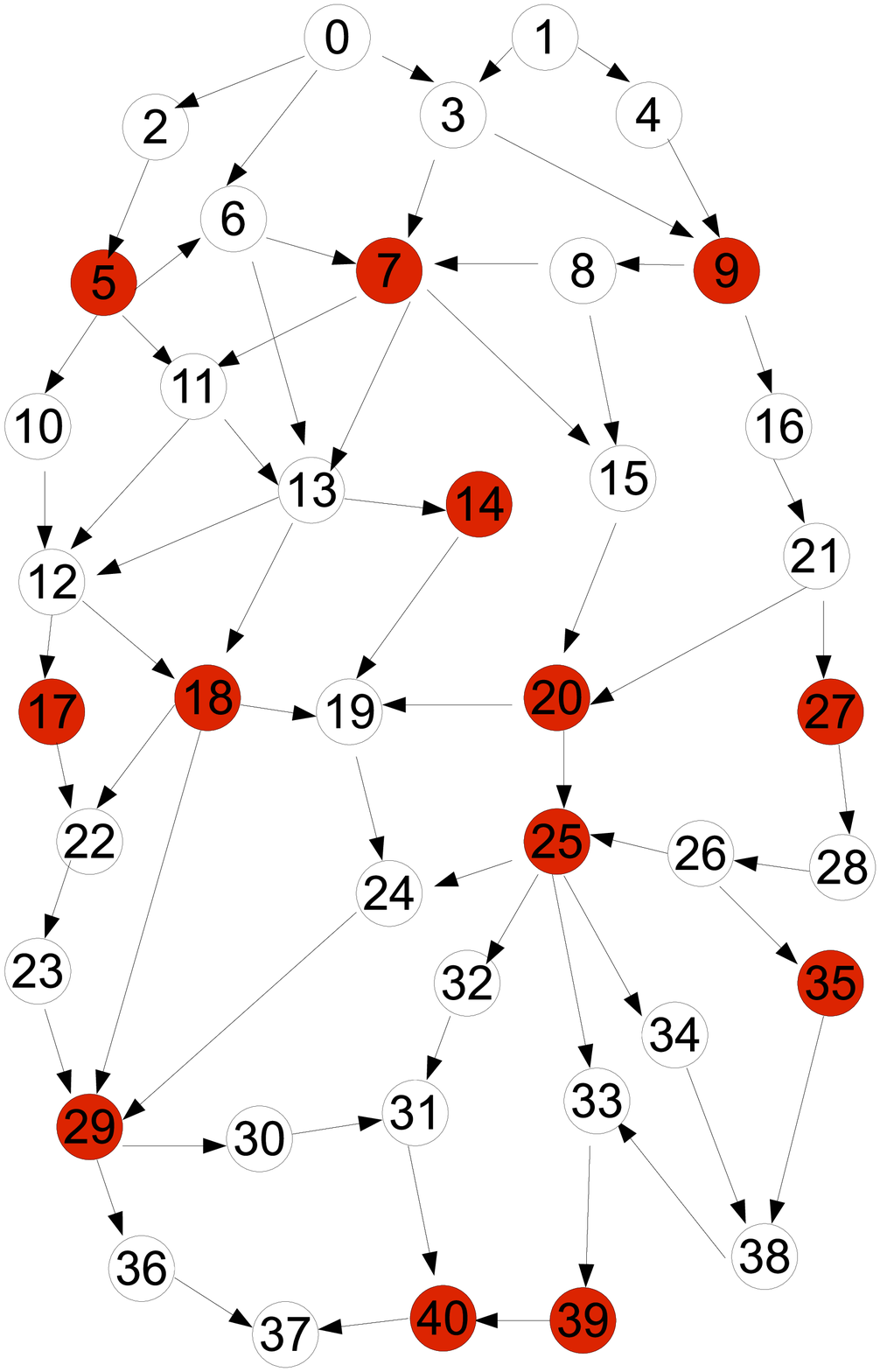}\label{graph}}
        \hspace{5mm}
        \subfigure[First Level Backbone $G_1$]{\includegraphics[scale=0.18]{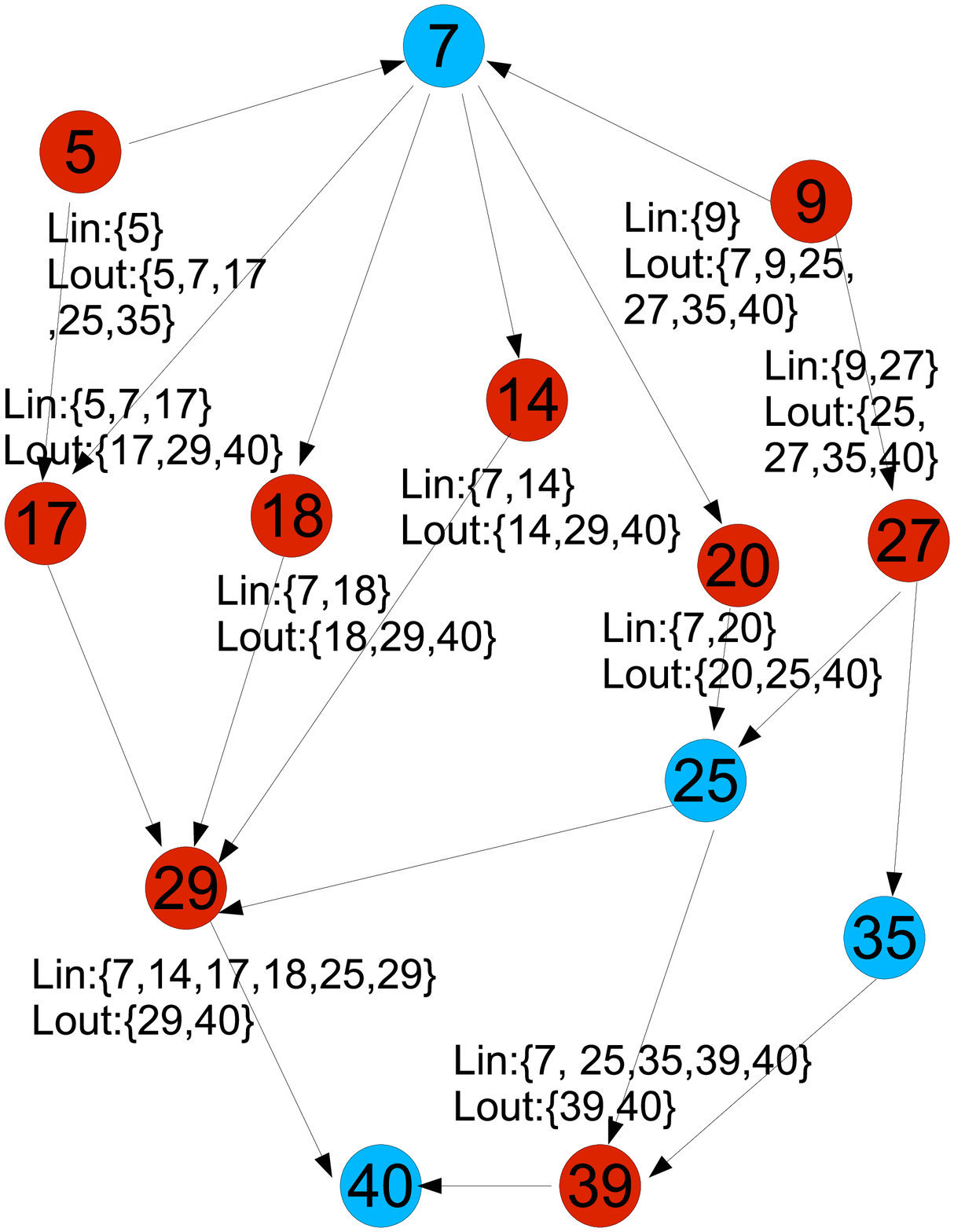}\label{level2}}
        \hspace{5mm}
        \subfigure[Second Level Backbone $G_2$]{\includegraphics[scale=0.13]{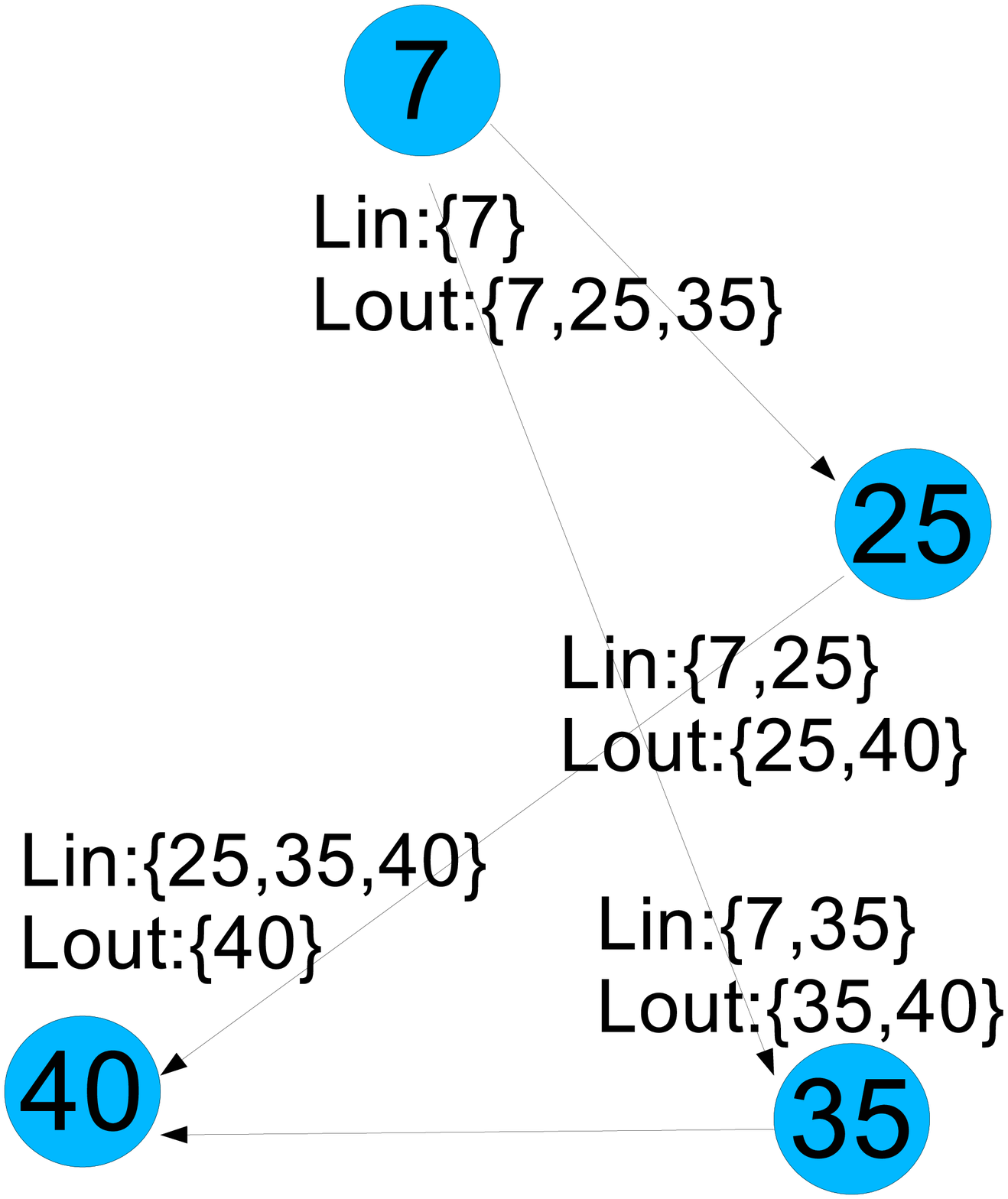}\label{core}}
        \hspace{5mm}
        \subfigure[Hop Labeling for $V_0$]{\includegraphics[scale=0.2]{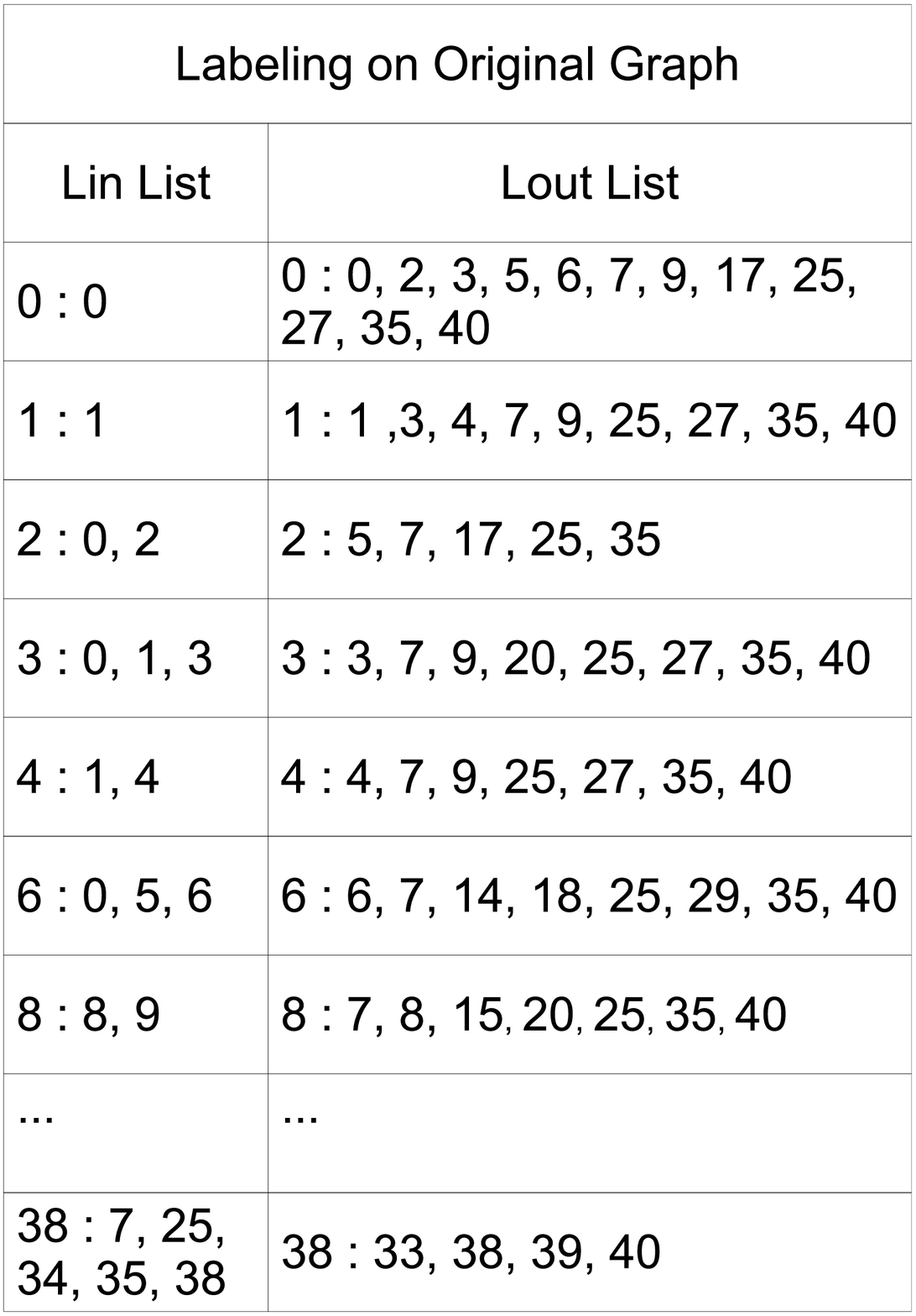}\label{labeling}}
    }
    \vspace*{-1.0ex}
    \caption{{\small Running Examples of Hierarchical-Labeling}}
\label{HLexample}
\end{figure*}

\begin{example}
As a simple example, let $V^\star$ be a vertex cover of $G$, i.e., at least one end of an edge in $E$ is in $V^\star$; and let $E^\star$ contain all edges $(u^\star,v^\star) \in V^\star \times V^\star$, such that $d(u^\star,v^\star) \leq 2$.
Then, $G^\star=(V^\star,E^\star)$ is one-side reachability backbone with $\epsilon=1$.
In Figure~\ref{HLexample}(b),  $G_1$ is the reachability backbone of graph $G_0$ (Figure~\ref{HLexample}(a)) for $\epsilon=2$.
\end{example}

The important property of the one-side reachability backbone is that {\em for any non-local pair $(u,v)$: $u \rightarrow v$ and $d(u,v) >\epsilon$, there always exists $u^\star \in V^\star$ and $v^\star \in V^\star$, such that $d(u,u^\star) \leq \epsilon$, $d(v^\star,v) \leq \epsilon$, and $u^\star \rightarrow v^\star$}. This property will serve as the key tool for recursively computing $L_{out}$ and $L_{in}$. In ~\cite{Jin:2012:SSR}, the authors develop the {\em FastCover} algorithm employing $\epsilon$-step BFS for each vertex for discovering the one-side reachability backbone. They also show that when $\epsilon=2$, the backbone can already be significantly reduced. To simplify our discussion, in this paper, we will focus on using the reachability backbone with $\epsilon=2$ though the approach can be applied to other locality threshold values.

Below, Subsection~\ref{hierachyalgorithm} presents the hierarchical decomposition of a DAG and the labeling algorithm using this DAG; Subsection~\ref{correctness} discusses the correctness of the labeling approach and its time complexity.

\vspace*{-1.0ex}
\subsection{Hierarchical  DAG Decomposition and Labeling Algorithm}
\label{hierachyalgorithm}

Let us start with the hierarchical DAG decomposition which is based  on the reachability backbone.

\bdefin({\bf  Hierarchical DAG Decomposition})
\label{DAGHD}
Given DAG $G=(V,E)$, a vertex hierarchy is defined as $V_0=V \supset V_1 \supset V_2 \supset \cdots \supset V_h$, with corresponding edge sets $E_0, E_1, E_2 \cdots E_h$, such that $G_i=(V_i,E_i)$ is the (one-side) reachability backbone of $G_{i-1}=(V_{i-1},E_{i-1})$, where $0<i\leq h$. The final graph $G_h=(V_h,E_h)$ is referred to as the core graph.
\edefin

Intuitively, the vertex hierarchy shows the relative importance of vertices in terms of reachability computation.
The lower level reachability computation can be resolved using the higher level vertices, but not the other way around.
In other words, the reachability (backbone) property is preserved through the vertex hierarchy.

\blemma
Assuming $u \in V_i, v \in V_i$, $u$ reaches $v$ in $G$ ($u \overset{G}{\longrightarrow}v$)  iff $u$ reaches $v$ in $G_i$ ($u \overset{G_i}{\longrightarrow}v$).
Furthermore, for any non-local vertex pairs $(u_i,v_i) \in V_i$, $d(u_i,v_i|G_i) >\epsilon$ (the distance in $G_i$), there always exists $u_{i+1} \in V_{i+1}$ and $v_{i+1} \in V_{i+1}$, such that $d(u_i,u_{i+1}|G_i) \leq \epsilon$, $d(v_{i+1},v_{i}|G_i) \leq \epsilon$, and $u_{i+1} \overset{G_{i+1}}{\longrightarrow} v_{i+1}$.
\elemma
\bproof
The first claim: assuming $u \in V_i, v \in V_i$, $u$ reaches $v$ in $G$ ($u \overset{G}{\longrightarrow}v$)  iff $u$ reaches $v$ in $G_i$ ($u \overset{G_i}{\longrightarrow}v$), can be proved by induction. 
The base case where $i=1$ is clearly true based on the reachability backbone definition (the reachability backbone will  preserve the reachability between vertices in the backbone as they appear in the original graph).  
Assuming this is true for all $i<k$, then it also holds to be true for $i=k$. This is because for any $u \in V_i, v \in V_i$, we must have $u \in V_{i-1}$ and $v \in V_{i-1}$. Based on the reachability backbone definition, we have $u \overset{G_{i-1}}{\longrightarrow}v$ iff $u \overset{G_{i-1}}{\longrightarrow}v$. Then based on the induction, we have $G$ ($u \overset{G_0=G}{\longrightarrow}v$)  iff $u$ reaches $v$ in $G_i$ ($u \overset{G_i}{\longrightarrow}v$).
The second claim directly follows the reachability definition. 
\eproof

\begin{example}
Figure~\ref{HLexample} shows a vertex hierarchy for DAG $G_0$ (a), where $V_1=\{5,7,9,\cdots,40\}$ (b) and $V_2=\{7,25,35,40\}$ (c). $G_1$ is the (one-side) reachability backbone of $G_0$ and $G_2$ is the corresponding (one-side) reachability backbone of $G_1$.
\end{example}

To utilize the hierarchical decomposition for labeling, let us further introduce a few notations related to the vertex hierarchy.
Each vertex $v$ is assigned to a unique {\em level}: $level(v)=i$ iff $v \in V_i \setminus V_{i+1}$, where $0 \leq i \leq h$ and $V_{h+1}=\emptyset$. (Later, we will show that each vertex is labeled at its corresponding level using $G_i$ and labels of vertices from higher levels). Assuming $v$ is at level $i$, i.e., $level(v)=i$, let $N_{out}^k(v|G_i)$ ($N_{in}^k(v|G_i)$) be the $v$'s {\em $k$-degree outgoing (incoming) neighborhood}, which includes all the vertices $v$ can reach (reaching $v$) within $k$ steps in $G_i$. Finally, for any vertex $v$ at level $i<h$,  its corresponding outgoing (incoming) backbone vertex set $\boldsymbol{\mathcal{B}}^\epsilon_{out}(v)$ ($\boldsymbol{\mathcal{B}}^\epsilon_{in}(v)$) is defined as:
{\small

\beqnarr
\boldsymbol{\mathcal{B}}^\epsilon_{out}(v)=\{u \in V_{i+1}| d(v,u|G_i) \leq \epsilon \mbox{ and  there is no other vertex } \nonumber \\ \mbox{   $x \in V_{i+1}$,  such that }
d(v,x|G_i) \leq \epsilon \wedge d(x,u|G_i) \leq \epsilon  (v \rightarrow x \rightarrow u) \}
\label{boutset} \\
\boldsymbol{\mathcal{B}}^\epsilon_{in}(v) = \{u \in V_{i+1}| d(u,v|G)_i) \leq \epsilon \mbox{ and there is no other vertex  } \nonumber \\ \mbox{  $y \in V_{i+1}$,   such that }
d(u,y|G_i) \leq \epsilon \wedge d(y,v|G_i) \leq \epsilon (u \rightarrow y \rightarrow v) \}
\label{binset}
\eeqnarr

}

Now, let us see how the labeling algorithm works given the hierarchical decomposition.
Contrary to the decomposition process which proceeds from the lower level to higher level (like peeling),
the labeling performs from the higher level to the lower level.
Specifically, it first labels the core graph $G_h$ and then iteratively labels the vertex at level $h-1$ to level $0$.

\noindent{\bf Labeling Core Graph $G_h$:}
Theoretically, the diameter of the core graph $G_h$ is no more than $\epsilon$ (the pairwise distance between any vertex pair in $G_h$ is no more than $\epsilon$), and thus no more reachability backbone is needed ($V_{h+1}=\emptyset$).
In this case, for a vertex $v \in V_h$ ($level(v)=h$),  the basic labeling can be as simple as follows:
\beqnarr
L_{out}(v) =N_{out}^{\lceil \epsilon/2 \rceil}(v|G_h); & L_{in}(v)=N_{in}^{\lceil \epsilon/2 \rceil}(v|G_h)
\label{corelabel}
\eeqnarr
The labeling is clearly complete for $G_h$ as any reachable pair is within distance $\epsilon$.
Alternatively, since the core graph is typically rather small, we can also employ the existing $2$-hop labeling algorithm~\cite{cohen2hop,hopiedbt} to perform the labeling for core graphs.
Given this, practically, the decomposition can be stopped when the vertex set $V_h$ is small enough (typically less than $10K$) instead of making its diameter less than or equal to $\epsilon$.

\noindent{\bf Labeling Vertices with Lower Level $i$ ($0\leq i<h$):}
After the core graph is labeled, the remaining vertices will be labeled in a level-wise fashion from higher level $h-1$ to lower level (until level $0$).
For each vertex $v$ at level $0\leq i<h$, assuming all vertices in the higher level ($>i$) have been labeled ($L_{out}$ and $L_{in}$), then the following simple rule can be utilized for labeling  $v$:
\beqnarr
L_{out}(v)  &=& N_{out}^{\lceil \epsilon/2 \rceil}(v|G_i) \cup (\bigcup_{u \in \boldsymbol{\mathcal{B}}_{out}^\epsilon(v|G_i)} L_{out}(u))
\label{lowerlevellabelLout}
\\ L_{in}(v)  &=&  N_{in}^{\lceil \epsilon/2 \rceil}(v|G_i) \cup (\bigcup_{u \in \boldsymbol{\mathcal{B}}_{in}^\epsilon(v|G_i)} L_{in}(u))
\label{lowerlevellabelLin}
\eeqnarr
Basically, the label of $L_{out}(v)$ ($L_{in}(v)$) at level $i$ consists of two parts: the outgoing (incoming) $\lceil \epsilon/2 \rceil$-degree neighbors of $v$ in $G_i$, and the labels from its corresponding outgoing (incoming) backbone vertex set $\boldsymbol{\mathcal{B}}_{out}^\epsilon(v|G_i)$ ($\boldsymbol{\mathcal{B}}_{in}^\epsilon(v|G_i)$).
In particular, if $\epsilon=2$ (the typical locality threshold), then each vertex $v$ basically records its direct outgoing (incoming) neighbors in $G_i$ and the labels from its backbone vertex set.%some essential vertices within two steps.

\begin{algorithm}
\caption{Hierachical-Labeling($G=(V,E)$)}
\label{alg:HL}
\begin{algorithmic}[1]
\STATE Perform Hierarchical Decomposition of $G$ based on Definition~\ref{DAGHD};
\STATE Labeling core graph $G_h$; 
\STATE $i \leftarrow h-1$; 
\WHILE{$i \geq 0$ \COMMENT{Labeling $V_i$ from higher level to lower}}
       \FORALL{$v \in V_i \setminus V_{i+1}$ \COMMENT{labeling each vertex specific for $V_i$}}
	 \STATE $L_{out}(v) \leftarrow N_{out}^{\lceil \epsilon/2 \rceil}(v|G_i) \cup (\bigcup_{u \in \boldsymbol{\mathcal{B}}_{out}^\epsilon(v|G_i)} L_{out}(u))$
	 \STATE $L_{in}(v) \leftarrow N_{in}^{\lceil \epsilon/2 \rceil}(v|G_i) \cup (\bigcup_{u \in \boldsymbol{\mathcal{B}}_{in}^\epsilon(v|G_i)} L_{in}(u))$
       \ENDFOR 
       \STATE $i \leftarrow i-1$; 
\ENDWHILE
\end{algorithmic}
\end{algorithm}

\noindent{\bf Overall Algorithm:}
Algorithm~\ref{alg:HL} sketches the complete Hierarchical-Labeling approach.
Basically, we first perform the recursive hierarchical DAG decomposition (Line $1$).
Then, the vertices at the core graph $G_h$ will be labeled either by Formula~\ref{corelabel} or using the existing $2$-hop labeling approach (Line $2$).
Finally, the while-loop performs the labeling from higher level $h-1$ to lower level $0$ iteratively (Lines $4$-$10$), where each vertex $v$ in the level $i$ (Lines $5-9$) will be labelled based on Formulas~\ref{lowerlevellabelLout} and ~\ref{lowerlevellabelLin}.
%Note that in the hierarchical decomposition, the {\em FastCover} algorithm~\cite{} will be used

\begin{example}
Figure~\ref{HLexample} illustrates the Hierarchical-Labeling process, where  Figure~\ref{HLexample}(c) shows the labeling of core graphs. Note that for simplicity, each vertex by default records itself in both $L_{in}$ and $L_{out}$, and $\epsilon=2$.
Figure~\ref{HLexample}(b) shows the labeling for vertices in $V_1$; and Table~\ref{HLexample}(c) illustrates the labeling of a few vertices in $V_0$.
Taking vertex $14$ for example: $L_{in}(14)$ records its direct incoming neighbors in $G_1$ $\{7,14\}$ (and itself), and other labels from the labels of its corresponding incoming backbone vertex set $\boldsymbol{\mathcal{B}}_{in}^\epsilon(14|G_1)=\{7\}$.
Thus, $L_{in}(14)=\{7,14\}$.
Now $L_{out}(14)$ records its direct outgoing neighbors $\{14,29\}$ and $L_{out}$ of vertex $40$ ($\boldsymbol{\mathcal{B}}_{out}^\epsilon(14|G_1)=\{40\}$).
\end{example}

%\input{Figures/HierarchyExample.tex}

%The pseudocode of HierarchyTwohop is as Algorithm 1.

%\input{Figures/HierarchyTwohop.tex}

%Then we give the pseudocode of Compute\_Lin Algorithm 2, the Compute\_Lout is just performing the same algorithm but on the reversed direction.

%\input{Figures/ComputeLin.tex}

\comment{
be discussed.  this section, we will present {\em Hierarchical-Labeling} approach.
In this approach, the hierarchical structure is produced by  a recursive reachability backbone approach, i.e., finding a reachability backbone $G^\star$ from the original graph $G$ and then apply the backbone extraction algorithm on $G^\star$. Recall that the reachability backbone is introduced by the latest SCARAB framework~\cite{Jin:2012:SSR} which aims to scale the existing reachability computation approaches. Here we applied it recursively to provide a hierarchical DAG decomposition. Given this, a fast labeling algorithm is designed to quickly compute $L_{in}$ and $L_{out}$ one vertex by one vertex in a level-wise fashion (from higher level to lower level).

The intuition of HierarchyTwohop is an extention of our previous work SCARAB, where we build one Backbone structure on top of the given graph to speed up the reachability computation. Here we push further, and not just build one Backbone structure, but iteratively use the generated Backbone as also given graph, and build another Backbone based on it. So that we generate one hierarchy with several levels of the the original graph and it will be finally scaled up into a "core" graph within several iterations. The number of iteration is normally less then 10, and the "core" graph contains only hundreds of vertices according to our experiment result.

We already know that for the top level, we use Twohop algorithm to assign labels. Here we formally describe the labeling algorithm for lower levels. For each vertex, it is only computed once when it's on the highest level that contains it. Consequently, when a vertex compute the Lin/Lout list, it only considers vertices that first appear on higher or the same levels.

And for computing Lin/Lout lists, starting from each vertex first appearing on this level we perform a BFS within epsilon distance. For all vertices it can reach, if reached vertex is a Backbone vertex, we add this vertex as well as it's corresponding Lin/Lout list into the Lin/Lout list of starting vertex. Otherwise we add only this vertex it self into Lin/Lout list of starting vertex.}

\vspace*{-1.0ex}
\subsection{Algorithm Correctness and Complexity}
\label{correctness}

In the following, we first prove the correctness of the Hierarchical-Labeling algorithm, that is, that it produces a complete labeling: for any vertex pair $(u,v)$, $u \rightarrow v$ iff $L_{out}(u) \cap L_{in}(v) \neq \emptyset$. We then discuss its time complexity.

\bthm
\label{HLcomplete}
The Hierarchical-Labeling approach (Algorithm~\ref{alg:HL}) produces a complete labeling for each vertex $v$ in graph $G$, such that for any vertex pair $(u,v)$: $u \rightarrow v \mbox{ iff  } L_{out}(u) \cap L_{in}(v) \neq \emptyset$.
\ethm
\bproof
We prove the correctness through induction: assuming Algorithm~\ref{alg:HL} produces the correct labeling for $V_{i+1}$, then it produces the correct labeling for $V_i$.
Basically if for any vertex pair $u^\star$ and $v^\star$ in $V_{i+1}$,  $u^\star \rightarrow v^\star$ iff $L_{out}(u^\star) \cap L_{in}(v^\star) \neq \emptyset$, then we would like to show that for any vertex pair $u$ and $v$ in $V_i$, this also holds.
To prove this, we consider four different cases for any $u$ and $v$ in $V_{i+1}$: 1) $u \in V_{i} \setminus V_{i+1}$ and $v \in V_{i} \setminus V_{i+1}$; 2) $u \in V_{i} \setminus V_{i+1}$ and $v \in V_{i+1}$; 3) $u \in V_{i+1}$ and  $v \in  V_{i} \setminus V_{i+1}$; and 4) $u \in V_{i+1}$ and $v \in V_{i+1}$.
Since case 4 trivially holds based on the reduction and cases 2 and 3 are symmetric, we will focus on proving cases 1 and 2.

\noindent{\bf Case 1 ($u \in V_{i} \setminus V_{i+1}$ and $v \in V_{i} \setminus V_{i+1}$)}:
We observe: 1) $u \rightarrow v$ with $d(u,v) \leq \epsilon$ (local pair) iff there is $x \in V_{i}$, such that $d(u,x) \leq \lceil \frac{\epsilon}{2} \rceil$ and $d(x,v) \leq \lceil \frac{\epsilon}{2} \rceil$, i.e., $N_{out}^{\lceil \epsilon/2 \rceil}(v|G_i) \cap N_{in}^{\lceil \epsilon/2 \rceil}(v|G_i) \neq \emptyset$;

and 2) $u \rightarrow v$ with $d(u,v) > \epsilon$ (non-local pair) \\ iff there are backbone vertices $u^\star, v^\star \in V_{i+1}$, such that $d(u,u^\star) \leq \epsilon$, $d(v^\star,v) \leq \epsilon$ and $u^\star \rightarrow v^\star$. That is, $L_{out}(u^\star) \cap L_{in}(v^\star) \neq \emptyset$ \\ iff there are $x \in \boldsymbol{\mathcal{B}}_{out}^\epsilon(u|G_i)$ and $y \in \boldsymbol{\mathcal{B}}_{in}^\epsilon(v|G_i)$, such that $x \rightarrow y$, i.e., $L_{out}(x) \cap L_{in}(y) \neq \emptyset$ (if there is $x \in V_{i+1}$, such that  $d(u,x) \leq \epsilon$ and $d(x,u^\star) \leq \epsilon$, then we can always use $x$ to replace $u^\star$ for the above claim; ($u^\star \rightarrow v^\star$ then $x \rightarrow v^\star$))
 \\ iff $(\bigcup_{u \in \boldsymbol{\mathcal{B}}_{out}^\epsilon(v|G_i)} L_{out}(u)) \cap (\bigcup_{u \in \boldsymbol{\mathcal{B}}_{in}^\epsilon(v|G_i)} L_{out}(u)) \neq \emptyset$.

\noindent{\bf Case 2 ($u \in V_{i} \setminus V_{i+1}$ and $v \in V_{i+1}$)}:
We observe 1) $u \rightarrow v$ with $d(u,v) \leq \epsilon$ (local pair) iff either $v \in \boldsymbol{\mathcal{B}}_{out}^\epsilon(u|G_i)$ ($v \in L_{out}(u)$ and $v \in L_{in}(v)$), or there is $x \in \boldsymbol{\mathcal{B}}_{out}^\epsilon(v|G_i)$, such that $x \rightarrow v$, i.e. $L_{out}(x) \cap L_{in}(v)\neq \emptyset$ \\
iff $(\bigcup_{u \in \boldsymbol{\mathcal{B}}_{out}^\epsilon(v|G_i)} L_{out}(u)) \cap (\bigcup_{u \in \boldsymbol{\mathcal{B}}_{in}^\epsilon(v|G_i)} L_{out}(u)) \neq \emptyset$;
and 2)  $u \rightarrow v$ with $d(u,v) >\epsilon$ (non-local pair) iff there exists $x$ such that $x \in \boldsymbol{\mathcal{B}}_{out}^\epsilon(v|G_i)$ and $x \rightarrow v$, i.e. $L_{out}(x) \cap L_{in}(v)\neq \emptyset$ \\
iff $(\bigcup_{u \in \boldsymbol{\mathcal{B}}_{out}^\epsilon(v|G_i)} L_{out}(u)) \cap (\bigcup_{u \in \boldsymbol{\mathcal{B}}_{in}^\epsilon(v|G_i)} L_{out}(u)) \neq \emptyset$.

Thus, in all cases, we have the correct labeling for any vertex pair $u$ and $v$ in $V_{i+1}$.
Now,  the core labeling is correct either based on the basic case where the graph diameter is no more than $\epsilon$ or based on the existing $2$-hop labeling approaches ~\cite{cohen2hop,hopiedbt}.
Together with the above induction rule, we have for any vertex pair in $V=V_0$, the label is complete and we thus prove the claim.
\eproof

\noindent{\bf Complexity Analysis:}
The computational complexity of Algorithm~\ref{alg:HL} comes from three components:
1) the hierarchical DAG decomposition, 2) the core graph labeling, and 3) the remaining vertex labeling for levels from $h-1$ to $0$.
For the first component, as we mentioned earlier, we can employ the {\em FastCover} algorithm~\cite{Jin:2012:SSR} iteratively to extract the reachability backbone vertices $V_i$ and their corresponding graph $G_i$. The {\em FastCover} algorithm is very efficient and to extract $G_{i+1}$ from $G_i$, it just needs to traverse the $\epsilon$ neighbors of each vertex in $G_{i+1}$. Its complexity is $O(\sum_{v \in V} \\|N_{out}^\epsilon(v|G_i)| log |N_{out}^\epsilon(v|G_i)|$ + $|E_{out}^\epsilon(v|G_i)|)$, where $E_{out}^\epsilon(v|G_i)$ is the set of edges $v$ can reach in $\epsilon$ steps.
Also, we note that in practice, the vertex set $V_i$ shrinks very quickly and after a few iterations ($5$ or $6$ typically for $\epsilon=2$), the number of backbone vertices is on the order of thousands (Section~\ref{expr}). We can also limit the total number of iterations, such as bounding $h$ to be $10$ and/or stop the decomposition when the $V_i$ is smaller than some limit such as $10K$.
For the second component, if the diameter is smaller than $\epsilon$ and Formula~\ref{corelabel} is employed, it also has a linear cost: $O(\sum_{v \in V} ( |N_{out}^\epsilon(v|G_h)| + |E_{out}^\epsilon(v|G_h)| +|N_{in}^\epsilon(v|G_h)| + |E_{in}^\epsilon(v|G_h)|))$. If we employ the existing $2$-hop labeling approach~\cite{cohen2hop,hopiedbt}, the cost can be $O(|V_h|^4$). However, since $|V_h|$ is rather small, the cost can be acceptable and in practice (Section~\ref{expr}), it is also quite efficient.
Finally, the cost to assign labels for all the remaining vertices is linear to their neighborhood cardinality and the labeling size of each vertex. It can be written as $O(\sum_{v \in V_i \setminus V_{i+1}} (|N_{out}^\epsilon(v|G_h)| + |E_{out}^\epsilon(v|G_h)| +|N_{in}^\epsilon(v|G_h)| + |E_{in}^\epsilon(v|G_h)|) + ML$, where $M$ is the maximal number of vertices in the backbone vertex set and $L$ is the maximal number of vertices in any $L_{in}$ or $L_{out}$.

We note that for large graphs, the last component typically dominates the total computational cost as we need to perform list merge (set-union) operations to generate $L_{out}$ and $L_{in}$ for each vertex. However, compared with the existing hop labeling approach, \\ Hierarchical-Labeling is significantly cheaper as there is no need for materializing transitive closure and the set-cover algorithm.  The experimental study (Section~\ref{expr}) finds that the labeling size produced by the Hierarchical-Labeling approach is comparable to that produced by the expensive set-cover based optimization.

%This confirms the hierarchical structure for defining the importance of hops.
%For random graphs with average vertex degree $d$ and assuming $\espilon=2$, then total time complexity can be written as

\comment{
Apparently our algorithm is based on one assumption that it's sufficient to compute each vertex only once when we processing the highest level that contains it. And to make sure the correctness, we proved the lemma below.

\blemma({\bf Sufficient labeling})
\label{lemma1}
$\forall x | x \in level_i$, there's no need to compute $L_{in}/L_{out}$ lists for x on $level_{j}$, where $j > i$.
\elemma

\bproof
To prove lemmar 1, first we prove it is correct for $L_{out}$ lists. let us consider three cases.

case 1. $BXB$.
u and v are both backbone vertices.
The labels generated from previous level will be sufficient to answer their reachability queries.

case 2. $\bar{B}XB$
u is a backbone vertex, but v is not.

When the distance between u and v is less or equal to epsilon, then when we perform BFS starting from v, we will definitely reach u and record u in v's Lin/Lout list.

When the distance between u and v is more than epsilon steps. According to the definition of Backbone, if u and v are reachable, there must be an x, such that u and x are reachable based on their labels, also x and v are reachable within epsilon steps.

case 3. $BX\bar{B}$.
Same as case2.

case 4. $\bar{B}X\bar{B}$
Neither u nor v are backbone vertices.

When the distance between u and v is less or equal to half of epsilon, then both u and v will reach each other during the process of BFS and record each other in their Lin/Lout lists.

When the distance between u and v is larger than half of epsilon but less or equal to epsilon, then if u and v are reachable, there must exist a backbone vertex x, such that both u and v can reach x within epsilon steps, and therefore record x in their Lin/Lout lists.

When the distance between u and x is larger than epsilon, if u and v are reachable, there must exists at least two backbone vertices x and x', such that x and x' are reachable according to their labels, u and x are reachable within epsilon steps, v and x are reachable within epsilon steps.
\eproof

%Figure 1 shows an example of how we generate the Hierarchy.

%Omitted for final submissionNote since the transitive reduction of a given graph is unique~\cite{AhoGU72}, the {\em canonical} backbone edge set  is also unique.

%We reduce the classical NP-hard problem, set-cover problem (SCP), to the decision version of this problem and the proof can be found in Appendix.
}

\vspace*{-2.0ex}
\section{Distribution Labeling}
\label{order}
%\vspace*{-1.0ex}

The Hierarchical-Labeling approach provides a fast alternative to produce a {\em complete} reachability oracle.
Its labeling is dependent on a reachability-based hierarchical decomposition and follows a process similar to the classical transitive closure computation~\cite{Simon88}, where the transitive closure of all incoming neighbors are merged to produce the new transitive closure. However, the potential issue is that when merging $L_{out}$ and $L_{in}$ of higher level vertices for the lower level vertices, this approach does not (and cannot) check whether any hop is redundant, i.e., their removal can still produce a complete labeling. Given the current framework, it is hard to evaluate the importance of each individual hop as they being cascaded into lower level vertices. Recall that for a vertex $v$, when computing its $L_{out}(v)$ and $L_{in}(v)$, its corresponding backbone vertex sets  ($\boldsymbol{\mathcal{B}}^\epsilon_{out}(v)$ and $\boldsymbol{\mathcal{B}}^\epsilon_{in}(v)$) only eliminate those redundant backbones if they can be linked through a local vertex (Formulas~\ref{boutset} and ~\ref{binset}).  Thus even if $u \in \boldsymbol{\mathcal{B}}^\epsilon_{out}(v)$, it may still be redundant as there is another vertex $u^\prime \in \boldsymbol{\mathcal{B}}^\epsilon_{out}(v)$ such that $u^\prime \rightarrow u$ (but $d(u^\prime,u)$ is large).
However, this issue is related to the difficulty of computing transitive reduction as mentioned earlier.

In light of these issues, we ponder the following: Can we perform labeling without the recursive hierarchical decomposition?
Can we explicitly confirm the ``power'' or ``importance'' of an individual hop as it is being added into $L_{out}$ and $L_{in}$? In this work, we provide positive answers to these questions and along the way, we discover a simple, fast, and elegant labeling algorithm, referred to as {\em Distribution-Labeling}: 1) the recursive hierarchical decomposition is replaced with a simple total order of vertices (the order criterion can be as simple as a basic function of vertex degree); 2) each hop is explicitly verified to be added into $L_{out}$ and $L_{in}$ only when it can cover some additional reachable pairs, i.e., it is non-redundant.
Surprisingly,  the labeling size produced by this approach is even smaller than the set-cover approach on all the available benchmarking graphs used in the recent reachability studies (Section~\ref{expr}).

In Subsection~\ref{hopcover}, we first introduce a simple yet fundamental observation of {\em hop-covering} (given a hop, what vertex pairs can it cover), which is the basis for the {\em Distribution-Labeling} algorithm; and Subsection~\ref{oracle}, we present the labeling algorithm and discuss its properties.

\begin{figure*}[t!]
%    \label{fig:scalabeloracleexample}
    \centering
    \mbox{
        \subfigure[Labeling for $Cov(13)$]{\includegraphics[scale=0.22]{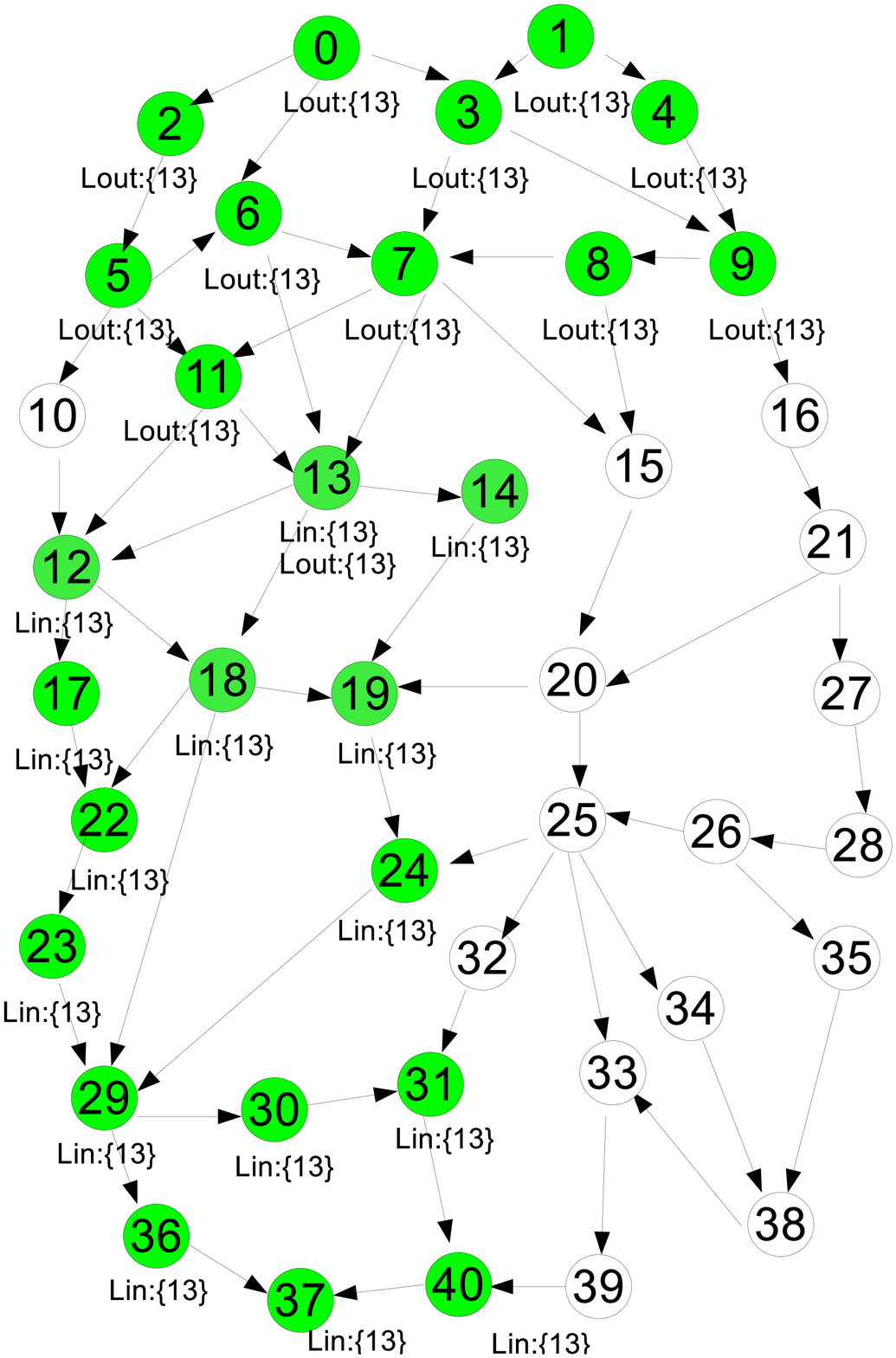}\label{label13}}
        \hspace{-1mm}
        \subfigure[Labeling for $Cov(\{13,7\})$]{\includegraphics[scale=0.22]{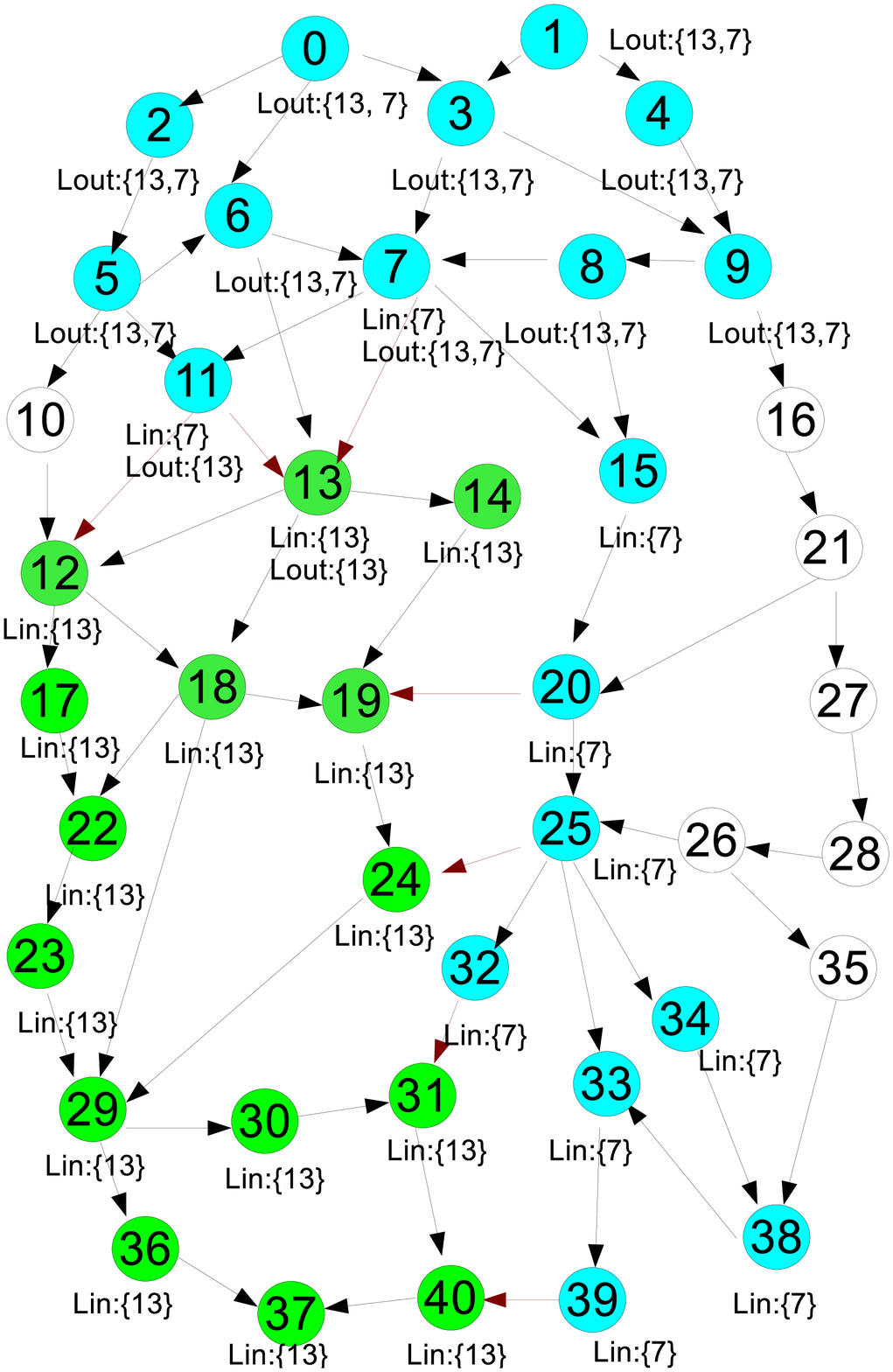}\label{label7}}
        \hspace{-1mm}
        \subfigure[Labeling for $Cov(\{13,7,25\})$]{\includegraphics[scale=0.22]{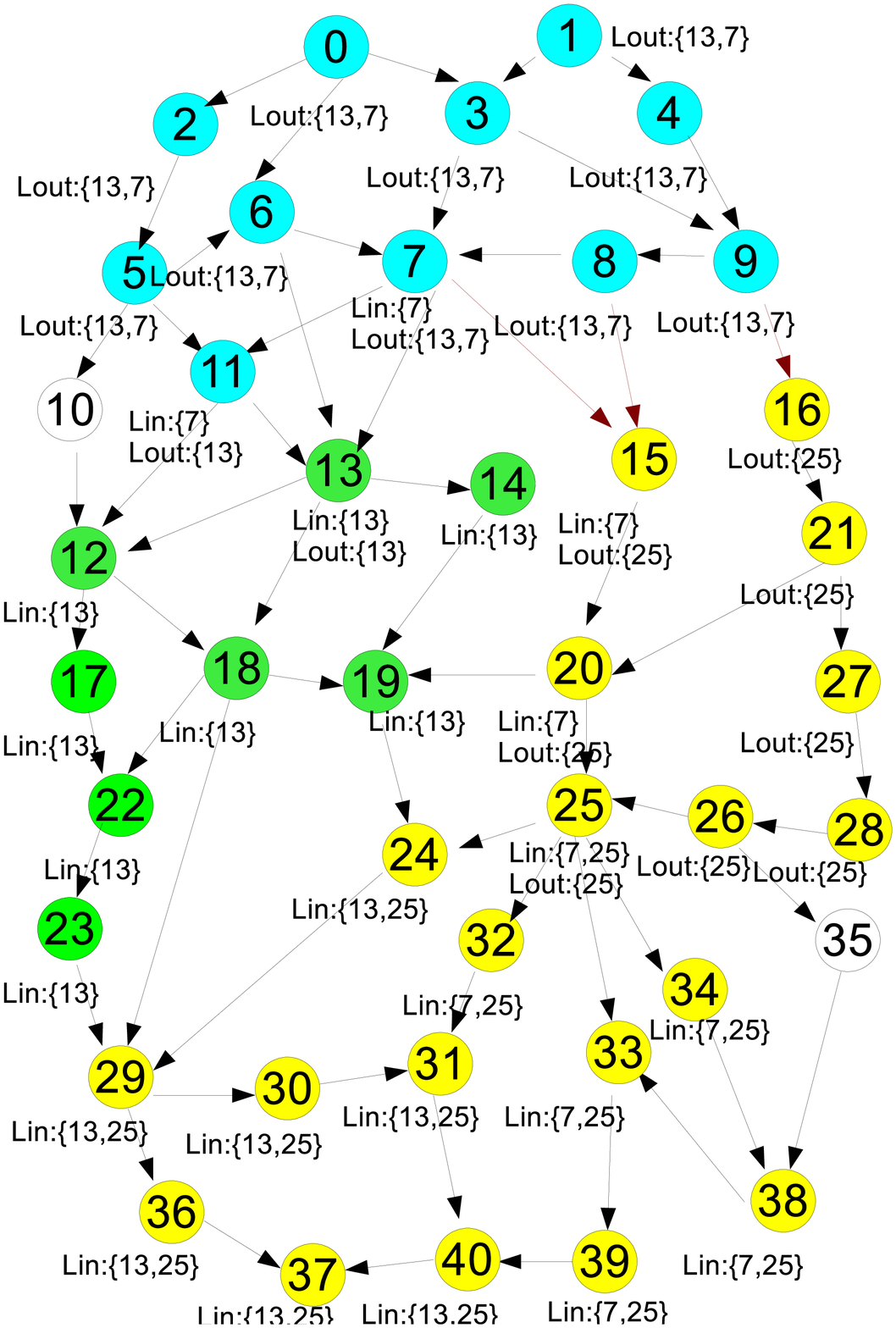}\label{label25}}
        \hspace{-1mm}
        \subfigure[ Basic Labeling]{\includegraphics[scale=0.22]{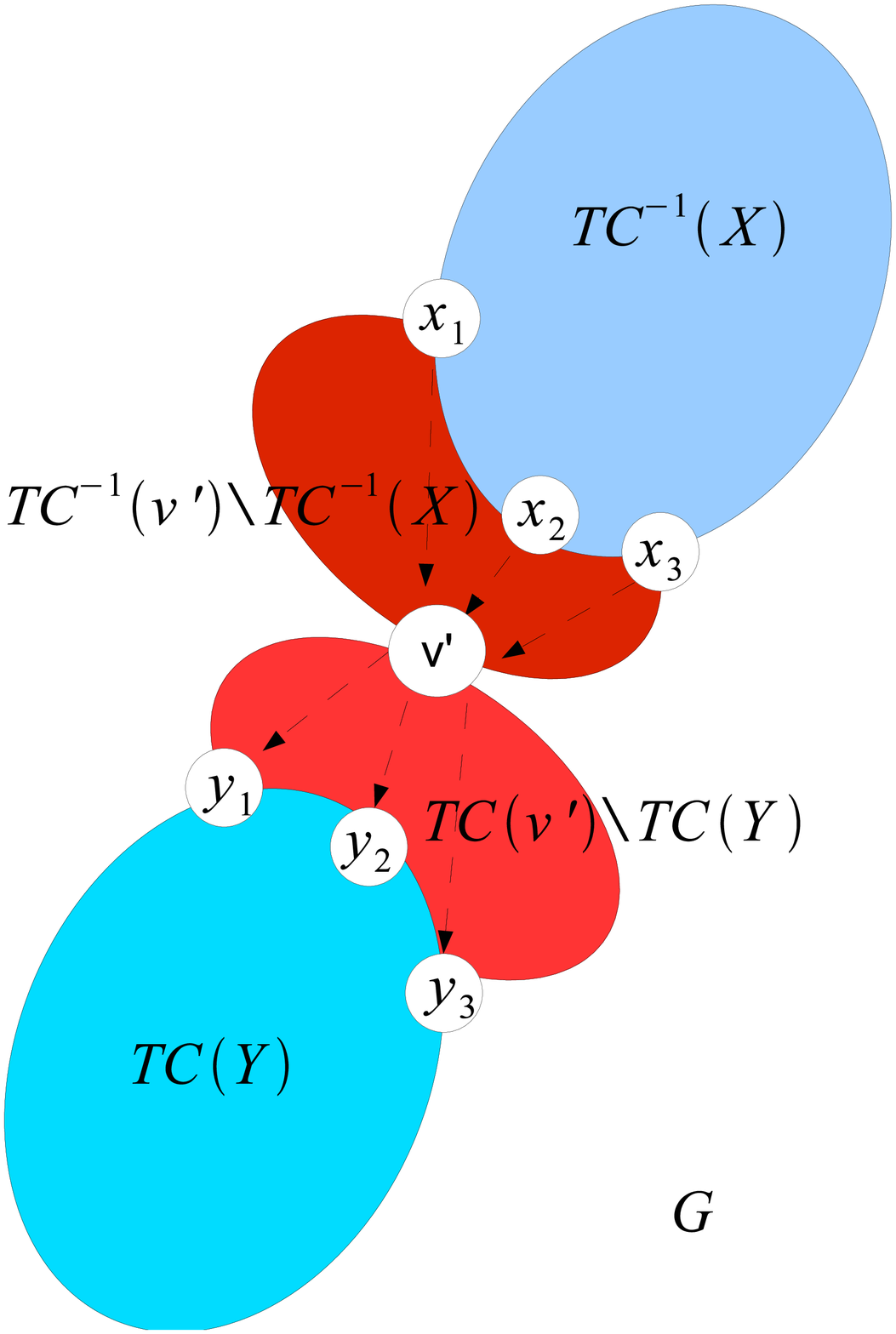}\label{TCcov}}
    }
    \vspace*{-3.0ex}
    \caption{{\small Running Example of Distribution-Labeling}}
\label{distributionlabelingexample}
\vspace*{-3.0ex}
\end{figure*}

\vspace*{-1.0ex}
\subsection{Hop Coverage and Labeling Basis}
\label{hopcover}

We first formally define the ``covering power'' of a hop and then study the relationship of two vertices in terms of their ``covering power''.

\bdefin({\bf Hop Coverage})
For vertex $v$, its {\bf coverage} $Cov(v)$ is defined as $TC^{-1}(v) \times TC(v)=\{(u,w): u \rightarrow v \mbox{ and } v \rightarrow w\}$. Note that $TC^{-1}(v)$ is the reverse transitive closure of $v$ which includes all the vertices reaching $v$. If for any pair in $(u,w) \in Cov(v)$, $L_{out}(u) \cap L_{in}(w) \neq \emptyset$, then we say $Cov(v)$ is covered by the labeling. We also say $Cov(v)$ can be covered by $v$ if each vertex $u$ reaching $v$ ($u \in TC^{-1}(v)$) has $v \in L_{out}(u)$ and each vertex $w$ being reached by $v$ has $v \in L_{in}(w)$ ($w \in TC(v)$).
\edefin

Given this, the labeling $L_{out}$ and $L_{in}$ is complete if it covers $Cov(V)=\cup_{v \in V} Cov(v)$, i.e., for any $(u,w) \in Cov(V)$,\\
To achieve a complete labeling, let us start with $Cov({v,v^\prime})=Cov(v) \cup  Cov(v^\prime)$. We study how to use only $v$ and $v^\prime$ to cover $Cov({v,v^\prime})$.
Specifically, we consider the following question: {\em assuming $v$ has been recorded by $L_{out}(u)$ for every $u \in TC^{-1}(v)$ and by $L_{in}(w)$ for every $w \in TC(v)$, then in order to cover the reachability pairs in $Cov({v,v^\prime})$ and only $v^\prime$ can serve as the hop, what vertices should record $v^\prime$ in their $L_{out}$ and $L_{in}$?}

To answer this question, we consider three cases: 1) $v$ and $v^\prime$ are incomparable, i.e., $v \nrightarrow v^\prime$ and $v \nleftarrow v^\prime$; 2)$v^\prime \rightarrow v$;  and 3) $v \rightarrow v^\prime$.
For the first case, the labeling is straightforward: each $u \in TC^{-1}(v^\prime)$ needs to record $v^\prime \in L_{out}(u)$ and each $w \in TC(v^\prime)$ needs to record $v^\prime \in L_{in}(u)$. Note that in the worst case, this is needed in order to recover pairs as $TC^{-1}(v^\prime) \times \{v^\prime\}$ and $\{v^\prime\} \times TC(v^\prime)$. For Cases 2 and 3, Lemma~\ref{keylemma} provides the answer.

\blemma
\label{keylemma}
Let $L_{out}(u)=\{v\}$ for every $u \in TC^{-1}(v)$ and  $L_{in}(w)=\{v\}$ for every $w \in TC(v)$.
If $v^\prime \rightarrow v$, then with $L_{out}(u)=\{v,v^\prime\}$ for $u \in TC^{-1}(v^\prime)$ and $L_{in}(w)=\{v^\prime\}$ for $w \in TC(v^\prime)\setminus TC(v)$ (other labels remain the same), $Cov(\{v,v^\prime\})$ is covered (using only hops $v$ and $v^\prime$).
If $v \rightarrow v^\prime$, then with $L_{out}(u)=\{v^\prime\}$ for $u \in TC^{-1}(v^\prime) \setminus TC^{-1}(v)$ and $L_{in}(w)=\{v,v^\prime\}$ for $w \in TC(v^\prime)$ (other labels remain the same), $Cov(\{v,v^\prime\})$ is covered (using only hops $v$ and $v^\prime$).
\elemma
\bproof
We will focus on proving the case where $v^\prime \rightarrow v$ as the case $v^\prime \rightarrow v$ is symmetric.
We first note that if $v^\prime \rightarrow v$, then $TC^{-1}(v^\prime) \subseteq TC^{-1}(v)$ and $TC(v^\prime) \supseteq TC(v)$.
Since $Cov(v)=TC{-1}(v) \times TC(v)$ is already covered by $v$, the uncovered pairs in $Cov(\{v,v^\prime\})$ can be written as
\beqnarr
Cov(\{v,v^\prime\}) \setminus Cov(v)= TC^{-1}(v^\prime) \times (TC(v^\prime) \setminus TC(v))  \nonumber
\eeqnarr
Given this, adding $v^\prime$ to $L_{out}(u)$ where $u \in TC^{-1}(v^\prime)$ and to $L_{in}(w)$ where $w \in TC(v^\prime) \setminus TC(v)$ can thus cover all the pairs in $Cov(\{v,v^\prime\})$.
\eproof

\begin{example}
Figure~\ref{distributionlabelingexample}(a)  shows the labeling for $Cov(13)$ and Figure~\ref{distributionlabelingexample}(b) shows that for $Cov(13,7)$ where $7 \rightarrow 13$.
In particular, $TC^{-1}(13)=TC^{-1}(7) \cup \{11\}$ and $TC(13) \subset TC(7)$.
For all $u \in TC^{-1}(7)$, we have $L_{out}(u)=\{7,13\}$ and for all $w \in L_{in}(7)\setminus L_{in}(13)$, we have $L_{in}(w)=\{7\}$.
\end{example}

Given Lemma~\ref{keylemma}, we consider the following general scenario: for a subset of hops $V_s \subset V$, assume $L_{out}$ and $L_{in}$ are correctly labeled using only hops in $V_s$ to cover $Cov(V_s)$. Now how can we cover $Cov(V_s \cup \{v^\prime\})$ by adding the only additional hop $v$ to $L_{in}$ and $L_{out}$? The following theorem provides the answer (Lemma~\ref{keylemma} can be considered a special case):

\bthm({\bf Basic Labeling})
\label{labelingtheorem}
Given a subset of hops $V_s \subset V$, let $L_{out}(u) \subseteq V_s$ and $L_{in}(u) \subseteq V_s$ be complete for covering $Cov(V_s)$, i.e., for any $(u,v) \in Cov(V_s)$, $L_{out}(u) \cap L_{in}(v) \neq \emptyset$.
To cover $Cov(V_s \cup \{v^\prime\})$ using additional hop $v^\prime$, the following labeling is complete:
\beqnarr
L_{out}(u) \leftarrow L_{out}(u) \cup \{v^\prime\},  u \in TC^{-1}(v^\prime) \setminus TC^{-1}(X)  \\
L_{in}(w) \leftarrow L_{in}(w) \cup \{v^\prime\}, w \in TC(v^\prime) \setminus TC(Y)   \ \ \ \ \
\eeqnarr
where $X= TC^{-1}(v^\prime) \cap V_s$ including all the vertices in $V_s$ reaching $v^\prime$ and $Y= TC(v^\prime) \cap V_s$ including all the vertices in $V_s$ that can be reached by $v^\prime$; $TC^{-1}(X)=\bigcup_{v \in X} TC^{-1}(v)$ and $TC(Y)=\bigcup_{v \in Y} TC(v)$.
\ethm
The theorem and its proof can be illustrated in Figure~\ref{distributionlabelingexample}(d). 

\bproof
We first observe the following relationships between the (reverse) transitive closure of $v^\prime$ and $X$, $Y$.
\beqnarr
TC^{-1}(v^\prime) \supseteq TC^{-1}(X); & &  TC(v^\prime) \subseteq TC(v), v \in X; \nonumber \\
TC(v^\prime) \supseteq TC(Y); & & TC^{-1}(v^\prime) \subseteq TC^{-1}(v), v \in Y; \nonumber
\eeqnarr
Thus, following the similar proof of Lemma~\ref{keylemma}, we can see that
{\small
\beqnarr
Cov(V_s \cup \{v^\prime\}) = Cov (V_s) \cup TC^{-1}(v^\prime) \times TC(v^\prime)  \nonumber \\
=Cov(V_s) \cup (TC^{-1}(v^\prime) \setminus TC^{-1}(X)) \cup TC^{-1}(X)) \nonumber \\
\times  ((TC(v^\prime) \setminus TC(Y)) \cup TC(Y)) \nonumber
\eeqnarr
\beqnarr
=Cov(V_s) \cup (TC^{-1}(v^\prime) \setminus TC^{-1}(X)) \times (TC(v^\prime) \setminus TC(Y)) \nonumber \\
\cup (TC^{-1}(v^\prime) \setminus TC^{-1}(X)) \times \bigcup_{v \in Y} TC(v^\prime)  \nonumber \\
\cup TC^{-1}(X) \times (TC(v^\prime) \setminus TC(Y)) \nonumber \\
\cup TC^{-1}(X) \times TC(Y) \nonumber
\eeqnarr
\beqnarr
=Cov(V_s) \cup (TC^{-1}(v^\prime) \setminus TC^{-1}(X)) \times (TC(v^\prime) \setminus TC(Y)), \nonumber \\
\mbox{since }(TC^{-1}(v^\prime) \setminus TC^{-1}(X)) \times TC(Y)  \subseteq Cov(V_s); \nonumber \\
TC^{-1}(X) \times (TC(v^\prime) \setminus TC(Y)) \subseteq Cov(V_s); \nonumber \\
TC^{-1}(X) \times TC(Y) \subseteq Cov(V_s) \nonumber
\eeqnarr
}
Thus, by adding $v^\prime$ to $L_{out}(u), u \in TC^{-1}(v^\prime) \setminus TC^{-1}(X)$ and to $L_{in}(w), w \in TC(v^\prime) \setminus TC(Y)$, the labeling will be complete to cover $Cov(V_s \cup \{v^\prime\})$.
\eproof

\begin{example}
Figure~\ref{distributionlabelingexample}(c) shows an example of $Cov(\{13,7\} \cup \{25\})$, where $X=\{13,7\}$ (both can reach $25$ and $Y=\emptyset$. Thus $25$ is added to $L_{out}(u), u \in TC^{-1}(25) \setminus (TC(13) \cup TC(7))$ and to $L_{in}(w), w \in TC(25)$.
\end{example}

%We also note that such a labeling is also non-redundant (adding $v^\prime$ to $L_{out}(u), u \in TC^{-1}(v^\prime) \setminus TC^{-1}(X)$ and to $L_{in}(w), w \in TC(v^\prime) \setminus TC(Y)$). This is because in the worst case this will be needed in order to cover $TC^{-1}(v^\prime) \setminus TC^{-1}(X) \times \{v^\prime\}$ and $\{v^\prime\} \times TC(v^\prime) \setminus TC(Y)$ as the labeling using $V_s$ cannot cover them.
%

\vspace*{-1.0ex}
\subsection{Distribution-Labeling Algorithm}
\label{oracle}

In the following, based on Lemma~\ref{keylemma} and Theorem~\ref{labelingtheorem}, we introduce the {\em Distribution-Labeling} algorithm, which will iteratively distribute each vertex $v$ to $L_{out}$ and $L_{in}$ of other vertices to cover $Cov(V_s \cup \{v\})$ ($V_s$ includes processed vertices).
Intuitively, it first selects a vertex $v_1$ and provides complete labeling for $Cov(v_1)$; then it selects the next vertex $v_2$, provides complete labeling for $Cov(\{v_1,v_2\})$ based on Lemma~\ref{keylemma}. It continues this process, at each iteration $i$ selecting a new vertex $v_i$ and producing the complete labeling for $Cov(V_s \cup \{v_i\})$ based on Theorem~\ref{labelingtheorem} where $V_s$ includes all the $i-1$ vertices which have been processed. The complete labeling will be produced when $V_s=V$.

Given this, two issues need to be resolved for this labeling process: 1) What should be the order in selecting vertices, and 2) How can we quickly compute $X$ (processed vertices which can reach the current vertex $v_i$) and $Y$ (processed vertices $v_i$ can reach), and identify $u \in TC^{-1}(v_i) \setminus TC^{-1}(X)$ and
$w \in TC(v_i) \setminus TC(Y)$.

\noindent{\bf Vertex Order:} The vertex order can be considered an extreme hierarchical decomposition, where each level contains only one vertex. Furthermore, the higher level the vertex, then the more important it is, the earlier it will be selected for covering, and the more vertices that are likely to record it in their $L_{out}$ and $L_{in}$ lists. There are many approaches for determining the vertex order. For instance, if following the set-cover framework, the vertex can be dynamically selected to be the {\em cheapest} in covering new pairs, i.e., $\frac{|TC^{-1}(v_i) \setminus TC^{-1}(X)|+|TC(v_i) \setminus TC(Y)|}{|Cov(V_s \cup \{v_i\}) \setminus Cov(V_s)|}$.
However, this is computationally expensive. 
We may also use $|Cov(v_i)|$ which measures the covering power of vertex $v$, but this still needs to compute transitive closure. 
In this study, we found the following rank function, $(|N_{out}(v)|+1) \times (|N_{in}(v)|+1)$, which measures the vertex pairs with distance no more than $2$ being covered by $v$, is a good candidate and can provides compact labeling. 
Indeed, we have used a similar criterion in ~\cite{Jin:2012:SSR} for selecting reachability backbone.
In the experimental evaluation (Section~\ref{expr}), we will also use this rank function for computing the distribution labeling. 

\noindent{\bf Labeling $L_{out}$ and $L_{in}$:} Given vertex $v_i$, we need to find (1) $u \in TC^{-1}(v_i) \setminus  TC^{-1}(X)$, i.e., the vertices reaching $v_i$ but not reaching by $v$ such that $v \rightarrow v_i$ and it has a higher order (already being processed); and (2) $w  \in TC(v_i) \setminus TC(Y)$, i.e., the vertices which can be reached by $v_i$ but cannot be reached by $v$ such that $v_i \rightarrow v$ and it has a higher order.
The straightforward way for solving (1) is to perform a reversed traversal and visit (expand) the vertices based on the reversed topological order; then once the visited vertex has a higher order then $v_i$, all its descendents (including itself) will be colored (flagged) to be be excluded from adding $v_i$ to $L_{out}$; thus $v_i$ will be added to $L_{out}$ for all uncolored vertices during the reverse traversal process. A similar ordered traversal process can be used for solving (2).
However, the (reverse) ordered traversal needs a priority queue which results in $O(|V|\log|V|+|E|)$ complexity at each iteration.
In this work, we utilize a more efficient approach that can effectively prune the traversal space and avoid the priority queue, which is illustrated in Algorithm~\ref{alg:DL}.

\begin{algorithm}
\caption{Distribution-Labeling(G=(V,E))}
\label{alg:DL}
\begin{algorithmic}[1]
\STATE Rank vertices in $G$ in certain order; 
\FORALL {$v_i \in V$ \COMMENT{from higher order to lower}}
%    \COMMENT{Step 1: Labeling $L_{out}(u)$ for $u \in TC^{-1}(v_i) \setminus \bigcup_{v \in X} TC^{-1}(v)$} \\
    \STATE Perform Reverse BFS starting from $v_i$, and for each vertex $u$ being visited: 
     \IF{$L_{out}(u) \cap L_{in}(v_i) \neq \emptyset$}
   \STATE Do not add $v_i$ to $L_{out}(u)$ nor expand $u$;
    \ELSE
          \STATE Add $v_i$ into $L_{out}(u)$ and expand $u$ in the reverse BFS;
     \ENDIF
%    \COMMENT{Step 2: Labeling $L_{in}(w)$ for $w \in TC(v_i) \setminus \bigcup_{v \in Y}TC(v^\prime)$} \\
    \STATE Perform BFS starting from $v_i$, and for each vertex $w$ being visited: 
    \IF{$L_{in}(w) \cap L_{out}(v_i) \neq \emptyset$}
         \STATE Do not add $v_i$ to $L_{in}(u)$ nor expand $w$;
    \ELSE
         \STATE Add $v_i$ into $L_{in}(w)$ and expand $w$ in the BFS;
    \ENDIF
\ENDFOR
\end{algorithmic}
\end{algorithm}

%As you can tell from the pseudocode that the main difference between intuitive and optimal ScalableOracle is the initialization of inhibit lists. Even though the overall time complexity of $Optimal\_LinLout$ in worst case is still $O(|V||E| $, the time cost of BFS for initializing inhibit lists is saved.

In Algorithm~\ref{alg:DL}, the iteration labeling process is sketched in the foreach loop (Lines $2$ to $15$).
The main procedure in computing $u \in TC^{-1}(v_i) \setminus TC^{-1}(X)$ for labeling $L_{out}$ is outlined in Lines $3-8$.
The main idea is that when visiting a vertex $u$, once $L_{out}(u) \cap L_{in}(v_i)$ is no longer empty, we can simply exclude $u$ and its descendents from consideration, i.e., $u \in TC^{-1}(X)$ (Lines $4-6$). Intuitively, this is because there exists a vertex $v$, such that $u \rightarrow v \rightarrow v_i$ and has order higher than $v_i$.
Similarly,  the procedure that computes $w \in TC(v_i) \setminus TC(Y)$ for labeling $L_{in}$ is outlined in Lines $9-14$.
Here, the condition $L_{in}(w) \cap L_{out}(v_i) \neq \emptyset$ is utilized to prune $w$ and its descendents to determine $L_{in}$ labeling.
Figure~\ref{distributionlabelingexample} illustrates the labeling process based on Algorithm~\ref{alg:DL} for the first three vertices $13$, $7$, and $25$.

\vspace*{-1.0ex}
\subsection{Completeness, Compactness, and Complexity}
In the following, we discuss the labeling completeness (correctness), compactness (non-redundancy), and time complexity.

\bthm({\bf Completenss})
\label{Completeness}
The {\em Distribution-Labeling} algorithm (Algorithm~\ref{alg:DL}) produces a complete $L_{out}$ and $L_{in}$ labeling, i.e., for any vertex pair $(u,v)$, $u \rightarrow v$ iff $L_{out}(u) \cap L_{in}(v) \neq \emptyset$.
\ethm
\bproof
$u \in TC^{-1}(v_i) \setminus  TC^{-1}(X)$ and 2) $w \in TC(v_i) \setminus TC(Y)$. They are symmetric and we will focus on 1).
Note that for $u \in TC^{-1}(v_i) \setminus  TC^{-1}(X)$, we need to exclude vertex $u^\prime$ such that $u^\prime \rightarrow v \rightarrow v_i$, where $v$ is already processed (has higher order than $v_i$).
Assuming the labeling is complete for $Cov(V_s)$, where $V_s=\{v_1,\cdots,v_{i-1}\}$, then $L_{out}(u^\prime) \cap L_{in}(v_i) \neq \emptyset$ (Line $4$).
If $u^\prime$ should be excluded, then its descendents from the BFS traversal will also be true and should also be excluded.
Furthermore, the reverse BFS can visit all vertices where this condition does not hold, i.e., $L_{out}(u) \cap L_{in}(v_i) = \emptyset$, and thus $u \in TC^{-1}(v_i) \setminus TC^{-1}(X)$.
\eproof

Theorem~\ref{Completeness} shows that the Distribution-Labeling algorithm is correct; but how compact is the labeling?
The following theorem shows an interesting {\em non-redundant} property of the produced labeling, i.e., no hop can be removed from $L_{in}$ or $L_{out}$ while preserving completeness. We note that this property has not been investigated before in the existing studies on reachability oracle and hop labeling~\cite{cohen2hop,hopiedbt,ChengYLWY06,DBLP:conf/edbt/ChengYLWY08, DBLP:conf/sigmod/JinXRF09,Cai:2010}.

\bthm({\bf Non-Redundancy})
The {\em Distribution-Labeling} algorithm (Algorithm~\ref{alg:DL}) produces a non-redundant $L_{out}$ and $L_{in}$ labeling, i.e., if any hop $h$ is removed from a $L_{out}$ or $L_{in}$ label set, then the labeling becomes incomplete.
\ethm
\bproof
We will show that 1) for any $u \in TC^{-1}(v_i) \setminus  TC^{-1}(X)$, $v_i$ cannot be removed from $L_{out}$; and 2) for any $w \in TC(v_i) \setminus TC(Y)$, $v_i$ cannot be removed from $L_{in}$. Note that when $v_i$ is being added to $L_{out}(u)$ and $L_{in}(w)$, it is non-redundant as the new labeling at least covers $(TC^{-1}(v_i) \setminus  TC^{-1}(X)) \times \{v_i\}$ and $\{v_i\} \times TC(v_i) \setminus TC(Y)$.

However, will any later processed vertex $v_j$, such that $i<j$, make $v_i$ redundant?
The answer is no because in this case (still focusing on the above covered pairs by $v_i$), $u \rightarrow v_j \rightarrow v_i$ (or $w \leftarrow v_j \leftarrow v_i$),
but the order of $v_i$ is higher than $v_j$ and $v_j$ will not be added $v_i$ into its $L_{out}$ or $L_{in}$.
In other words, for any vertex pair in $(TC^{-1}(v_i) \setminus  TC^{-1}(X)) \times \{v_i\}$ or $\{v_i\} \times TC(v_i) \setminus TC(Y)$, $v_i$ is the only hop linking these pairs, i.e., $L_{out}(u) \cap L_{in}(v_i)= \{v_i\}$ and $L_{out}(v_i) \cap L_{out}(u) =\{v_i\}$.
Thus, $v_i$ is non-redundant for all the vertices recording it as label, i.e., $L_{out}(u), u \in TC^{-1}(v_i) \setminus  TC^{-1}(X)$ and $L_{in}(w),  w \in TC(v_i) \setminus TC(Y)$.
\eproof

As we discussed earlier,  Hierarchical-Labeling does not have this property; we can see this through counter-examples. For instance, in Figure~\ref{HLexample}(b), $17$ is redundant for $L_{out}(5)$. However, to remove these cases, the transitive reduction would have to be performed, which is expensive.
Furthermore, whether the labels produced by the existing set-cover based approach~\cite{cohen2hop} are redundant or not remains an open question though we conjecture they might be redundant.

\noindent{\bf Time Complexity:}
The worst case computational complexity of Algorithm~\ref{alg:DL} can be written as $O(|V|(|V|+|E|)L)$, where $L$ is the maximal labeling size. However, the conditions in Line $4$ and $10$ can significantly prune the search space, and $L$ is typically rather small, the {\em Distribution-Labeling} can perform labeling very efficiently. In the experimental study (Section~\ref{expr}), we will show Algorithm~\ref{alg:DL} is on average more than an order of magnitude faster than the existing hop labeling and has comparable or faster labeling time than the state-of-the-art reachability indexing approaches on large graphs. Its labeling size is also small and surprisingly, even smaller than the greedy set-cover based labeling approaches in most of the cases. This may be an evidence that the labeling of the existing set-cover based approach ~\cite{cohen2hop} is redundant.

\comment{

The intuitive idea is that we can rank all the vertices based on their capability of covering reach-ability information, which we measure as $(indegree + 1)*(outdegree + 1)$ descendingly. And since there's no epsilon restriction, we cannot use the same labeling strategy, we proposed even a simpler one.

First we introduce a concept called "inhibit list", which is the set of vertices that a vertex can reach by BFS, and has higher rank than the current vertex. The formal definitaion is as below.

\bdefin({\bf Inhibit List})
\label{def:inhibit}
Given a DAG $G=(V,E)$, $\forall v \in V$, $inhibit_{in}(v) = \{x \in V| rank(x) \leq rank(v) \& v \rightarrow x\}$.
\edefin

Then we propose a lemma that dramatically reduces the potential label size.

\blemma({\bf labeling by rank})
\label{lemma2}
    During the BFS starting from u, for any visited vertex $v_t$ , if $v_t \in inhibit_{in}(u)$, or $\exists x \in L_{in}(v_t)|x \in inhibit_{in}(u)$, then there's no need to add u to $L_{in}(v_t)$.
\elemma

\bproof
To prove this lemma, first we prove it is correct for $L_{out}$ lists. let us consider two cases.

First perform BFS from $u$ and visit $v_t$, where $rank(v_t) \leq rank(u)$. In this case, because there is a path from $u$ to $v_t$, $TC^{-1}(u) \subseteq TC^{-1}(v_t)$. $\forall x \in TC^{-1}(u), v_t \in L_{out}(x)$. So there is no need to add $u$ into $L_{out}(x)$ anymore.

Second perform BFS from u and visit $v_t'$, where $\exists x \in L_{in}(v_t') | x \in inhibit(u)$. So there must exist a path from $u$ to $v_t'$ through $x$, and $rank(x) \leq rank(u)$, so $TC^{-1}(u) \subseteq TC^{-1}(x)$. There is no need to add $u$ into $L_{out}(x)$ anymore.

In the same way, this lemma is proved correct with $L_{in}$.
\eproof

The pseudocode is as Algorithm 3.
\begin{algorithm}
\caption{Intuitive\_ScalableOracle\_IN(G)}
\label{alg:cl}
\begin{algorithmic}[1]
\STATE Rank vertices in G by $(indegree + 1)*(outdegree + 1)$ in descending order;
\STATE Initial inhibit list;
\FORALL {$u_i \in G$}
    \STATE Perform BFS starting from $u_i$, and record all visited vertices with rank higher than $u_i$ in $inhibit_{in}(u_i)$;
    \STATE Perform another BFS starting from $u_i$, and for each visited vertex $v_j$;
    \IF{$v_j \in inhibit_{in}(u_i)$ or $\exists x L_{in}(v_j)|x \in inhibit_{in}(u_i)$}
        \STATE Do not add $u_i$ into $L_{in}(v_j)$, nor extend BFS from $v_j$;
    \ELSE
        \STATE Add $u_i$ into $L_{in}(v_j)$, extend BFS from $v_j$;
    \ENDIF
    \STATE reset inhibit;
\ENDFOR
\end{algorithmic}
\end{algorithm}

Ranking vertices takes $O(|V|)$ time. Then actual labeling takes $O(|V||E|)$ time in worst case. Therefore, the overall time complexity of $Intuitive\_LinLout$ is $O(|V||E|)$.

\vspace*{-1.0ex}
\subsection{ScalableOracle optimization}
\label{so}
The intuitive ScalableOracle can reduce the redundanct lables efficiently. However, in order to initial the inhibit lists, for each vertex, two times of BFS is required, which is too expensive to be practical. To address this problem, we propose an optimal algorithm, that use the elements in $L_{out}$ list to initial $inhibit_{in}$, and the $L_{in}$ list to initial $inhibit_{out}$

\blemma({\bf initial inhibit})
\label{lemma3}
It is sufficient to initial $inhibit_{out}$ with elements in $L_{in}$ list and initial $inhibit_{in}$ with elements in $L_{out}$ list to reduce redundant labels.
\elemma

Define a special vertex on each path as $in\_minimal(u^{'})$, such that in all paths thought out-going edges starting from $u$, $u \rightarrow u^{'}$ and $rank(u^{'}) \leq rank(u)$. Additionally it is not the case that $\exists u^{''} such that u \rightarrow u^{''} \rightarrow u^{'}$ and $rank(u^{''}) \leq rank(u^{'}) \leq rank(u)$

Similarly we define $out\_minimal(u^{'})$, that in all paths thought in-coming edges starting from $u$, $u^{'} \rightarrow u$ and $rank(u) \leq rank(u^{'})$. Additionally it is not the case that $\exists u^{''} such that u^{'} \rightarrow u^{''} \rightarrow u$ and $rank(u^{''}) \leq rank(u^{'}) \leq rank(u)$

Since we compute $L_{in}$ and $L_{out}$ lists in the ascending order of rank, when we start initializing the inhibit list of $u$, $L_{out}$ must contains all $in\_minimal(u^{'})$ and $L_{in}$ must contains all $out\_minimal(u^{'})$.

\bproof
To prove lemmar 2, first we prove it is correct for $inhibit_{out}$ lists.

The reason is that, for any vertex $u^{''}$ such that $u^{''} \rightarrow u$, $rank(u^{''}) \leq rank(u)$, and $u^{''} \notin L_{out}(u)$.

There must exist a $out\_minimal(u^{'})$, such that $u^{''} \rightarrow u^{'} \rightarrow u$ and $rank(u^{'}) \leq rank(u^{''})$.

Because $u^{''} \rightarrow u^{'}$, $TC^{-1}(u^{'}) \supseteq TC^{-1}(u^{''})$.

Also $rank(u^{'}) \leq rank(u^{''})$, therefore $\forall x| x\in TC^{-1}(u^{''}), u^{'} \in L_{out}(x)$ and $u^{''} \notin L_{out}(x)$.

Consequentially there's no need to add $u^{''}$ into $inhibit_{out}(u)$.

In the same way, lemmar 2 is proved correct with $inhibit_{in}$.
\eproof

\comment{
\begin{figure}
\centering
\begin{tabular}{c c}
\psfig{figure=Figures/soSOidx1.eps,width=1.6in, height=2in} & \raisebox{0.5in}{\psfig{figure=Figures/soSOidx2.eps,width=0.8in, height=0.8in}}  \\
(a) Labeling start from 13  & (b) Labeling start from 7
\psfig{figure=Figures/soSOidx3.eps,width=1.6in, height=2in} & \raisebox{0.5in}{\psfig{figure=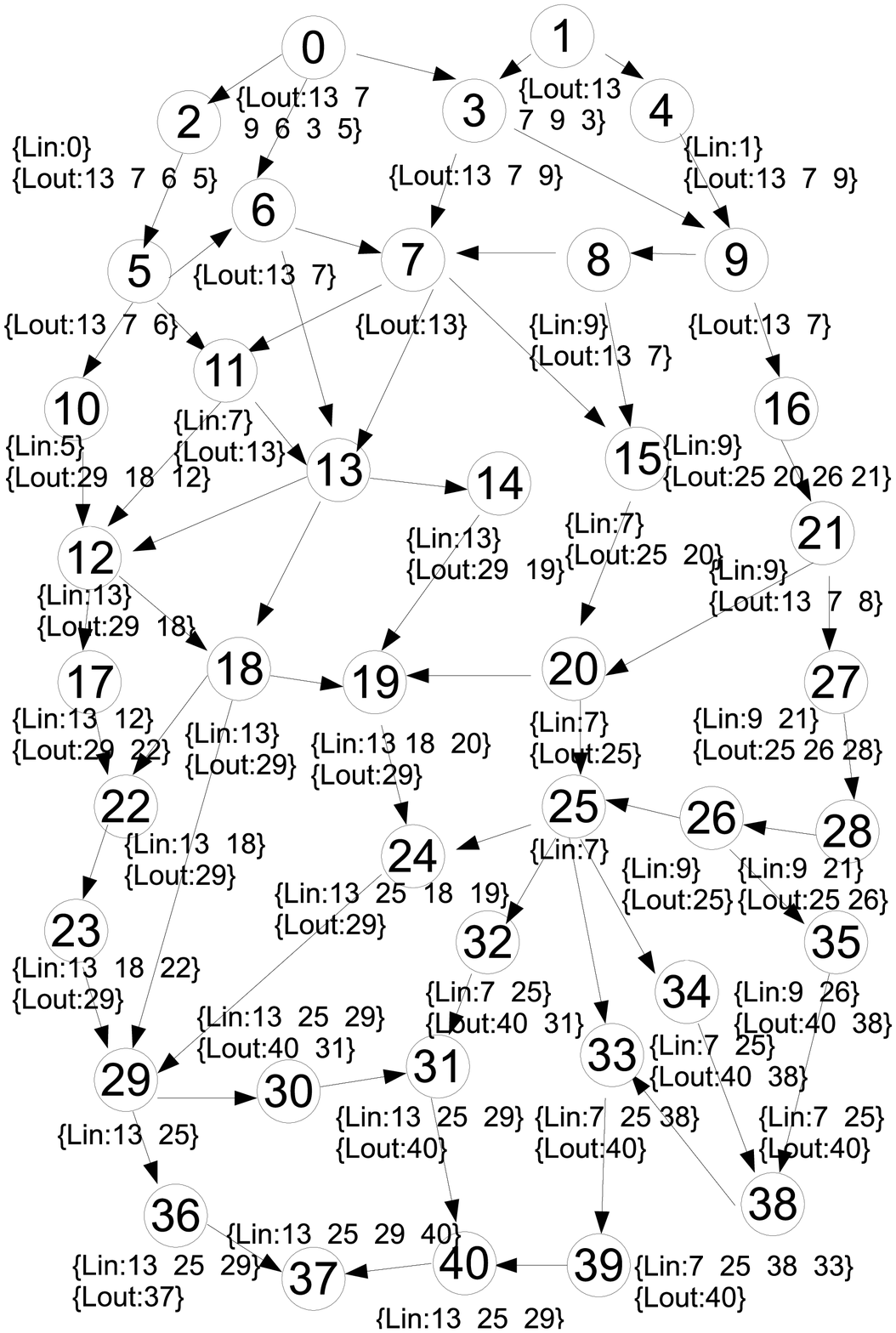,width=0.8in, height=0.8in}}  \\
(c) Labeling start from 25  & (d) Final Graph
\end {tabular}
\vspace*{-2.0ex}
\caption{Example of Optimal Scalableoracle labeling}
\label{fig:runningexamples}
\vspace*{-3.0ex}
\end{figure}}

The pseudocode is as Algorithm 4.}

\vspace*{-2.0ex}
\section{Experimental Evaluation}
\label{expr}

\begin{table}
\begin{center}
{\scriptsize
\begin{tabular}{|l|c|c|l|c|c|}
\hline
\multicolumn{3}{|c|}{Small Real Graph} & \multicolumn{3}{c|}{Large Real Graph} \\ \hline
Dataset & $|V|$  & $|E|$ & Dataset & $|V|$  & $|E|$ \\ \hline

agrocyc	&	12,684 	&	13,408 	&	citeseer	&	693,947 	&	312,282 	\\ \hline
amaze	&	3,710 	&	3,600 	&	citeseerx	&	6,540,399 	&	15,011,259 	\\ \hline
anthra	&	12,499 	&	13,104 	&	cit-Patents	&	3,774,768 	&	16,518,947 	\\ \hline
arxiv	&	21,608 	&	116,805 	&	email	&	231,000 	&	223,004 	\\ \hline
ecoo	&	12,620 	&	13,350 	&	go\_uniprot	&	6,967,956 	&	34,770,235 	\\ \hline
hpycyc	&	4,771 	&	5,859 	&	lj	&	971,232 	&	1,024,140 	\\ \hline
human	&	38,811 	&	39,576 	&	mapped\_100K	&	2,658,702 	&	2,660,628 	\\ \hline
kegg	&	3,617 	&	3,908 	&	mapped\_1M	&	9,387,448 	&	9,440,404 	\\ \hline
mtbrv	&	9,602 	&	10,245 	&	uniprotenc\_100m	&	16,087,295 	&	16,087,293 	\\ \hline
nasa	&	5,605 	&	7,735 	&	uniprotenc\_150m	&	25,037,600 	&	25,037,598 	\\ \hline
p2p	&	48,438 	&	55,349 	&	uniprotenc\_22m	&	1,595,444 	&	1,595,442 	\\ \hline
reactome	&	901 	&	846 	&	web	&	371,764 	&	517,805 	\\ \hline
vchocyc	&	9,491 	&	10,143 	&	wiki	&	2,281,879 	&	2,311,570 	\\ \hline
xmark	&	6,080 	&	7,028 	&	 	&	  	&	  	\\ \hline				

\end{tabular}
}
\end{center}
\vspace*{-4.0ex}
\caption{Real datasets} \label{realdata}
\vspace*{-2.0ex}
\end{table}

In this section, we empirically evaluate the {\em Hierarchical-Labeling} and {\em Distribution-Labeling} labeling algorithms against the state-of-the-art reachability computation approaches on a range of real graphs which have been widely used for studying reachability~\cite{yildirim:grail,vanSchaik:2011,DBLP:journals/tods/JinRXW11,Jin:2012:SSR,Cheng:2012:KYS}.
In particular, we are interested in the following questions ( in terms of the query efficiency, construction cost, and index (labeling) size) : 1) how do the reachability oracle approaches perform compared with the transitive closure compression and online search approaches? 2) how do these two approaches perform compared with the existing $2$-hop approaches assuming the later one can complete the labeling? 3) How do these two methods (Hierarchical-Labeling and  Distribution-Labeling) compare with one another?

\vspace*{-1.0ex}
\subsection{Experimental Setup}
To answer these questions, we evaluate the {\em Hierarchical-Labeling (HL)} and {\em Distribution-Labeling (DL)} labeling algorithms against the state-of-the-art reachability computation approaches:

\noindent{\bf 1)} {\em PathTree} ({\bf PT})~\cite{DBLP:conf/sigmod/JinXRW08}, an improved version of Agrawal's tree-interval method~\cite{SIGMOD:AgrawalBJ:1989};

\noindent{\bf 2)} {\em Nuutila's Interval} ({\bf INT}) ~\cite{Nuutila:1995}, a transitive closure compression method, recently demonstrated to be one of the fastest reachability computation methods~\cite{vanSchaik:2011};

\noindent{\bf 3)} {\em PAWH-8} ({\bf PW8})~\cite{vanSchaik:2011}, the latest bit-vector compression method for transitive closure compression)~\cite{vanSchaik:2011} and {\em PWAH-8} is its best variant~\cite{vanSchaik:2011}.

\noindent{\bf 4)} {\em K-Reach} ({\bf KR}) ~\cite{Cheng:2012:KYS}, a latest vertex-cover based approach for general reachability computation, i.e., determine whether two vertices are within distance $k$. Here $k$ is set to be the total number of vertices in the graph for the basic reachability.

\noindent{\bf 5)} {\em GRAIL} ({\bf GL}) ~\cite{yildirim:grail}, a scalable reachability indexing approach using random DFS labeling (the number of intervals is set at $5$, as suggested by authors).

\noindent{\bf 6)} {\em 2HOP} ({\bf 2HOP}) ~\cite{cohen2hop}, Cohen {\em et al.}'s  $2$-hop labeling approach;

Here, Path-Tree (1), Interval (2), and PAWH-8(3) are the state-of-the-art transitive closure compression approaches; K-Reach (4) is the latest general reachability approach and has been shown to be very capable in dealing with basic reachability~\cite{Cheng:2012:KYS} (it can also be considered as transitive closure compression as it materializes the transitive closure for the vertex-cover, a subset of vertices);  GRAIL (5) is the state-of-the-art online search approach; and 2HOP (6) is the existing set-cover based hop labeling approach. In addition, we also include the latest  {\em SCARAB} method~\cite{Jin:2012:SSR} for scaling PathTree and speeding up GRAIL, referred to as {\em PATH-TREE$^\ast$} ({\bf PT$^\ast$}) and {\em GRAIL$^\ast$} ({\bf GL$^\ast$}), respectively. The locality parameter $\epsilon$ is set at $2$ for SCARAB. 
We also add the comparison with the latest reachability labeling {\em TF-label} ({\bf TF})~\cite{DBLP:conf/sigmod/ChengHWF13}, and the latest distance labeling method {\em Pruned Landmark} ({\bf PL})~\cite{DBLP:conf/sigmod/AkibaIY13}. 

All the methods (including source code) except 2HOP and PL are either downloaded from authors' websites or provided by the authors directly. We have implemented 2HOP, Hierarchical-Labeling (HL), Distribution-Labeling (DL), PL, and 2HOP has been improved with several fast heuristics~\cite{hopiedbt,DBLP:conf/sigmod/JinXRF09} to speed up its construction time.
All these algorithms are implemented in C++ based on the Standard Template Library (STL). We also downloaded and tested {\em IS-labeling} (for distance computation)~\cite{DBLP:journals/corr/abs-1211-2367}. However, its query performance is at least 2 to 3 orders magnitude slower than the reachability methods; we omit reporting its results for the space limitation.  

In the experiments, we focus on reporting the three key measures for reachability computation: query time, construction time, and index size.
For the query time, similar to the latest SCARAB work~\cite{Jin:2012:SSR}, both {\em equal} and {\em random} reachability query workload are used. The equal query workload has about $50\%$ positive (reachable pairs) and about $50\%$ negative (unreachable pairs) queries. Positive queries are generated by sampling the transitive closure.
Also the query time is the running time of a total of $100,000$ reachability queries.

%For methods combined with {\bf SCARAB}~\cite{}, only indexing time is counted towards construction time, the time of discovering the Backbone graph is not included. Additionally, the index size of a {\bf SCARAB} combined method only consider the label size of given method on the backbone. The index size of the reachability backbone and the original graph size are not included.

All experiments are performed on a Linux 2.6.32 machine with Intel Xeon 2.67GHz CPU and 32GB RAM.

%Indeed, the first two parts of total index size can be used to better answer the second question we mentioned.

%\input{Figures/SmallRealDatasets.tex}
%\input{Figures/LargeRealDatasets.tex}

\comment{
\subsection{Datasets}

We briefly describe the real datasets and synthetic datasets, both small and large in the following.

Small real datasets (15 datasets): agrocyc, amaze, anthra, ecoo, human, kegg, mtbrv, nasa, vchocyc, xmark (used in pathtree sigmod and tods paper).

Large real datasets (8 datasets): cit-patents, citeseer, citeseerx, go-uniprot, uniprot22m, uniprot100m, uniprot150m, mapped100m (used in Grail paper).

Large synthetic datasets: we test all approaches on scale-free DAG synthetic datasets.
We generate the scale-free DAGs whose indegree follows power-law distribution, with the number of vertices from $1M$ to $10M$, while fixing the edge density as $2$.

}

\begin{table*}[!ht]
\centering
%\resizebox{6.5in}{!} {
{\small
\begin{tabular}{|l|r|r|r|r|r|r|r|r|r|r|r|r|}
\hline
\multicolumn{1}{|l|}{Dataset} &
\multicolumn{1}{l|}{GL} &
\multicolumn{1}{l|}{GL$^\ast$} &
\multicolumn{1}{l|}{PT} &
\multicolumn{1}{l|}{PT$^\ast$} &
\multicolumn{1}{l|}{KR} & \multicolumn{1}{l|}{PW8} & \multicolumn{1}{l|}{INT}&\multicolumn{1}{l|}{2HOP} & \multicolumn{1}{l|}{PL} & \multicolumn{1}{l|}{TF} & \multicolumn{1}{l|}{HL} & \multicolumn{1}{l|}{DL}
\\ \hline

agrocyc	&	189.8 	&	115.5 	&	1.1 	&	13.2 	&	1.5 	&	7.9 	&	2.6 	&	3.8 	&	122.6 	&	19.4 	&	4.3 	&	2.6 	\\ \hline
amaze	&	343.5 	&	28.8 	&	1.2 	&	8.7 	&	1.4 	&	3.5 	&	3.1 	&	3.0 	&	277.7 	&	2.6 	&	2.9 	&	2.5 	\\ \hline
anthra	&	124.2 	&	92.6 	&	1.3 	&	13.2 	&	1.5 	&	7.7 	&	2.6 	&	3.8 	&	77.2 	&	16.5 	&	3.9 	&	2.6 	\\ \hline
anthra	&	282.5 	&	214.2 	&	2.8 	&	15.2 	&	---	&	11.4 	&	3.7 	&	7.0 	&	90.9 	&	17.3 	&	11.8 	&	3.4 	\\ \hline
ecoo	&	122.1 	&	125.7 	&	1.1 	&	13.5 	&	1.5 	&	7.7 	&	2.7 	&	3.9 	&	74.9 	&	18.6 	&	4.4 	&	1.4 	\\ \hline
hpycyc	&	87.8 	&	25.0 	&	1.1 	&	13.8 	&	1.5 	&	8.6 	&	1.5 	&	3.8 	&	264.1 	&	14.2 	&	4.0 	&	1.3 	\\ \hline
human	&	185.4 	&	89.5 	&	1.2 	&	16.5 	&	1.8 	&	4.4 	&	3.2 	&	3.6 	&	81.4 	&	19.3 	&	2.5 	&	1.7 	\\ \hline
kegg	&	272.1 	&	44.9 	&	1.2 	&	16.5 	&	1.5 	&	4.8 	&	2.6 	&	3.2 	&	73.6 	&	2.7 	&	3.4 	&	2.0 	\\ \hline
mtbrv	&	115.4 	&	118.0 	&	1.1 	&	13.6 	&	1.5 	&	7.2 	&	2.6 	&	3.9 	&	64.7 	&	4.5 	&	5.1 	&	2.2 	\\ \hline
nasa	&	135.8 	&	126.8 	&	1.4 	&	21.5 	&	2.2 	&	18.4 	&	4.9 	&	4.4 	&	65.2 	&	3.4 	&	4.1 	&	3.6 	\\ \hline
p2p	&	1117.8 	&	308.4 	&	1.2 	&	6.5 	&	---	&	3.3 	&	1.7 	&	2.0 	&	48.3 	&	3.2 	&	3.6 	&	2.1 	\\ \hline
reactome\hspace{7 mm}	&	111.8 	&	22.8 	&	1.1 	&	11.6 	&	1.8 	&	12.1 	&	3.0 	&	3.1 	&	256.0 	&	2.0 	&	2.9 	&	2.1 	\\ \hline
vchocyc	&	107.6 	&	97.0 	&	1.0 	&	13.6 	&	1.5 	&	7.9 	&	2.6 	&	3.7 	&	65.0 	&	16.9 	&	3.8 	&	2.5 	\\ \hline
xmark	&	134.7 	&	255.9 	&	1.4 	&	21.6 	&	1.9 	&	35.8 	&	4.9 	&	5.9 	&	72.7 	&	5.8 	&	6.5 	&	4.4 	\\ \hline

\end{tabular}
}
%}
\vspace{-2.0ex}
\caption{Query Time (ms) Based on Equal Query of Small Real Datasets}
\label{smallrealdataEqualquery}
\vspace*{-1.0ex}
\end{table*}

\begin{table*}[!ht]
\centering
%\resizebox{7.0in}{!} {
{\small
\begin{tabular}{|l|r|r|r|r|r|r|r|r|r|r|r|r|}
\hline
\multicolumn{1}{|l|}{Dataset} &
\multicolumn{1}{l|}{GL} &
\multicolumn{1}{l|}{GL$^\ast$} &
\multicolumn{1}{l|}{PT} &
\multicolumn{1}{l|}{PT$^\ast$} &
\multicolumn{1}{l|}{KR} & \multicolumn{1}{l|}{PW8} & \multicolumn{1}{l|}{INT}&\multicolumn{1}{l|}{2HOP} & \multicolumn{1}{l|}{PL} & \multicolumn{1}{l|}{TF} & \multicolumn{1}{l|}{HL} & \multicolumn{1}{l|}{DL}
\\ \hline

agrocyc	&	29.0 	&	11.8 	&	1.4 	&	12.7 	&	1.2 	&	2.6 	&	2.2 	&	4.2 	&	133.9 	&	4.6 	&	4.4 	&	4.3 	\\ \hline
amaze	&	502.0 	&	23.8 	&	1.8 	&	15.0 	&	2.3 	&	3.7 	&	4.4 	&	4.0 	&	196.8 	&	2.9 	&	4.1 	&	4.2 	\\ \hline
anthra	&	29.6 	&	11.8 	&	1.4 	&	12.3 	&	1.2 	&	1.3 	&	2.2 	&	4.2 	&	369.9 	&	4.2 	&	4.4 	&	4.3 	\\ \hline
anthra	&	1209.4 	&	170.2 	&	4.2 	&	14.0 	&	---	&	14.7 	&	5.7 	&	4.2 	&	184.4 	&	22.3 	&	8.9 	&	5.5 	\\ \hline
ecoo	&	30.4 	&	11.8 	&	1.3 	&	12.3 	&	1.4 	&	2.5 	&	2.2 	&	4.2 	&	113.2 	&	4.2 	&	4.5 	&	4.3 	\\ \hline
hpycyc	&	28.8 	&	11.8 	&	1.3 	&	11.8 	&	2.0 	&	3.1 	&	2.3 	&	3.8 	&	132.6 	&	4.5 	&	4.0 	&	4.2 	\\ \hline
human	&	33.6 	&	16.7 	&	1.5 	&	16.1 	&	1.2 	&	1.5 	&	1.4 	&	2.7 	&	262.8 	&	3.1 	&	4.8 	&	2.9 	\\ \hline
kegg	&	616.6 	&	33.1 	&	1.9 	&	16.3 	&	2.6 	&	4.5 	&	4.6 	&	4.1 	&	111.4 	&	3.2 	&	4.4 	&	4.4 	\\ \hline
mtbrv	&	28.9 	&	12.6 	&	1.3 	&	12.8 	&	1.1 	&	1.4 	&	2.2 	&	3.9 	&	107.6 	&	2.7 	&	4.1 	&	4.2 	\\ \hline
nasa	&	28.2 	&	23.2 	&	1.7 	&	15.0 	&	2.7 	&	8.0 	&	5.3 	&	4.8 	&	111.9 	&	2.4 	&	5.3 	&	6.3 	\\ \hline
p2p	&	2279.6 	&	5.7 	&	1.6 	&	4.2 	&	---	&	1.9 	&	1.7 	&	2.5 	&	122.6 	&	2.8 	&	2.6 	&	3.2 	\\ \hline
reactome\hspace{7 mm}	&	32.9 	&	11.4 	&	3.0 	&	9.9 	&	3.0 	&	7.9 	&	3.4 	&	3.5 	&	101.4 	&	1.4 	&	3.4 	&	4.1 	\\ \hline
vchocyc	&	29.6 	&	12.2 	&	1.4 	&	12.5 	&	1.4 	&	2.6 	&	2.2 	&	4.0 	&	124.4 	&	4.3 	&	3.8 	&	4.2 	\\ \hline
xmark	&	63.7 	&	26.7 	&	1.7 	&	14.4 	&	2.3 	&	11.2 	&	4.5 	&	5.5 	&	126.6 	&	3.8 	&	5.4 	&	6.4 	\\ \hline

\end{tabular}
}
%}
\vspace{-2.0ex}
\caption{Query Time (ms) Based on Random Query of Small Real Datasets}
\label{smallrealdataRandomquery}
\vspace*{-1.0ex}
\end{table*}

\begin{table*}[!ht]
\centering
%\resizebox{7.0in}{!} {
{\small
\begin{tabular}{|l|r|r|r|r|r|r|r|r|r|r|r|r|}
\hline
\multicolumn{1}{|l|}{Dataset} &
\multicolumn{1}{l|}{GL} &
\multicolumn{1}{l|}{GL$^\ast$} &
\multicolumn{1}{l|}{PT} &
\multicolumn{1}{l|}{PT$^\ast$} &
\multicolumn{1}{l|}{KR} & \multicolumn{1}{l|}{PW8} & \multicolumn{1}{l|}{INT}&\multicolumn{1}{l|}{2HOP} & \multicolumn{1}{l|}{PL} & \multicolumn{1}{l|}{TF} & \multicolumn{1}{l|}{HL} & \multicolumn{1}{l|}{DL}
\\ \hline

agrocyc	&	22.6 	&	2.3 	&	128.1 	&	44.0 	&	284.5 	&	5.0 	&	3.7 	&	245.6 	&	18.8 	&	28.5 	&	120.8 	&	12.6 	\\ \hline
amaze	&	7.4 	&	0.9 	&	357.4 	&	18.4 	&	330.4 	&	4.5 	&	3.2 	&	2672.2 	&	5.4 	&	3.9 	&	43.2 	&	4.1 	\\ \hline
anthra	&	14.1 	&	2.3 	&	88.2 	&	41.4 	&	246.3 	&	4.1 	&	2.9 	&	241.0 	&	18.2 	&	44.9 	&	89.4 	&	12.4 	\\ \hline
anthra	&	45.0 	&	22.9 	&	131873.0 	&	8615.7 	&	---	&	100.6 	&	90.8 	&	145332.0 	&	384.0 	&	759.5 	&	618.6 	&	38.2 	\\ \hline
ecoo	&	12.8 	&	2.2 	&	94.5 	&	43.1 	&	282.1 	&	5.0 	&	3.7 	&	254.6 	&	18.2 	&	30.9 	&	92.2 	&	12.5 	\\ \hline
hpycyc	&	4.7 	&	0.9 	&	39.0 	&	11.2 	&	223.6 	&	2.7 	&	1.8 	&	199.2 	&	8.0 	&	16.5 	&	41.5 	&	5.2 	\\ \hline
human	&	71.2 	&	4.7 	&	298.2 	&	75.5 	&	296.5 	&	5.3 	&	4.1 	&	417.5 	&	43.8 	&	96.6 	&	155.2 	&	37.4 	\\ \hline
kegg	&	4.1 	&	1.0 	&	436.0 	&	19.4 	&	411.8 	&	5.6 	&	2.2 	&	2878.0 	&	6.2 	&	4.3 	&	48.3 	&	2.4 	\\ \hline
mtbrv	&	9.4 	&	1.7 	&	71.7 	&	30.9 	&	249.3 	&	2.2 	&	3.0 	&	208.3 	&	14.6 	&	14.6 	&	115.3 	&	9.8 	\\ \hline
nasa	&	10.2 	&	2.2 	&	49.4 	&	21.1 	&	1637.5 	&	9.8 	&	6.1 	&	835.9 	&	11.4 	&	10.5 	&	143.5 	&	8.9 	\\ \hline
p2p	&	69.8 	&	10.8 	&	2942.6 	&	432.6 	&	---	&	14.8 	&	25.6 	&	21618.9 	&	76.2 	&	45.9 	&	564.5 	&	51.1 	\\ \hline
reactome\hspace{7 mm}	&	1.2 	&	0.7 	&	8.3 	&	6.9 	&	35.1 	&	1.2 	&	0.8 	&	161.4 	&	0.6 	&	1.2 	&	25.4 	&	1.0 	\\ \hline
vchocyc	&	9.3 	&	1.7 	&	70.3 	&	30.9 	&	260.1 	&	4.4 	&	3.2 	&	224.3 	&	1.2 	&	45.4 	&	65.3 	&	9.5 	\\ \hline
xmark	&	11.1 	&	1.8 	&	109.3 	&	22.0 	&	806.2 	&	10.4 	&	5.9 	&	1557.0 	&	18.4 	&	19.5 	&	53.2 	&	8.7 	\\ \hline

\end{tabular}
%}
}
\vspace{-2.0ex}
\caption{Construction Time (ms) of Small Real Datasets}
\label{smallrealdataConstruction}
\vspace*{-1.0ex}
\end{table*}

\vspace*{-1.0ex}
\subsection{Experimental Results}
\label{small}

In the following, we report the experimental results on small graphs first and then on large graphs.
These graphs have been widely used for studying reachability computation ~\cite{Wang06,ChengYLWY06,DBLP:conf/sigmod/JinXRW08,DBLP:conf/sigmod/JinXRF09,Zhu:2009, yildirim:grail,Cai:2010,vanSchaik:2011,DBLP:journals/tods/JinRXW11,Cheng:2012:KYS,Jin:2012:SSR}.
In Table~\ref{realdata}, the first three columns give the names, number of vertices and number of edges for the coalesced DAGs derived from each original graph. The last three columns give similar information for large real graphs.

\noindent{\bf Small Graphs:}
Table~\ref{smallrealdataEqualquery} reports the query times of  the reachability oracle approaches (2HOP, Hierarchical-Labeling (HL), and Distribution-Labeling (DL)) against the state-of-the-art transitive closure compression approaches (PWAH-8,  INTERVAL, PATH-TREE, K-REACH), and online search (GRAIL), as well as some of their  SCARAB counterparts, including GRAIL$^\ast$ ({\bf GL$^\ast$}) and PATH-TREE$^\ast$ ({\bf PT$^\ast$})using the {\em equal} query load. We also compare them with the latest reachability labeling {\em TF-label} ({\bf TF})~\cite{DBLP:conf/sigmod/ChengHWF13}, and the latest distance labeling method {\em Pruned Landmark} ({\bf PL})~\cite{DBLP:conf/sigmod/AkibaIY13}. 

Table~\ref{smallrealdataRandomquery} reports the query time using the {\em random} query load.

\begin{figure*}[!ht]
%\resizebox{7.0in}{!} {
	\begin{minipage}[l]{6.5in}
	\centering
	\psfig{figure=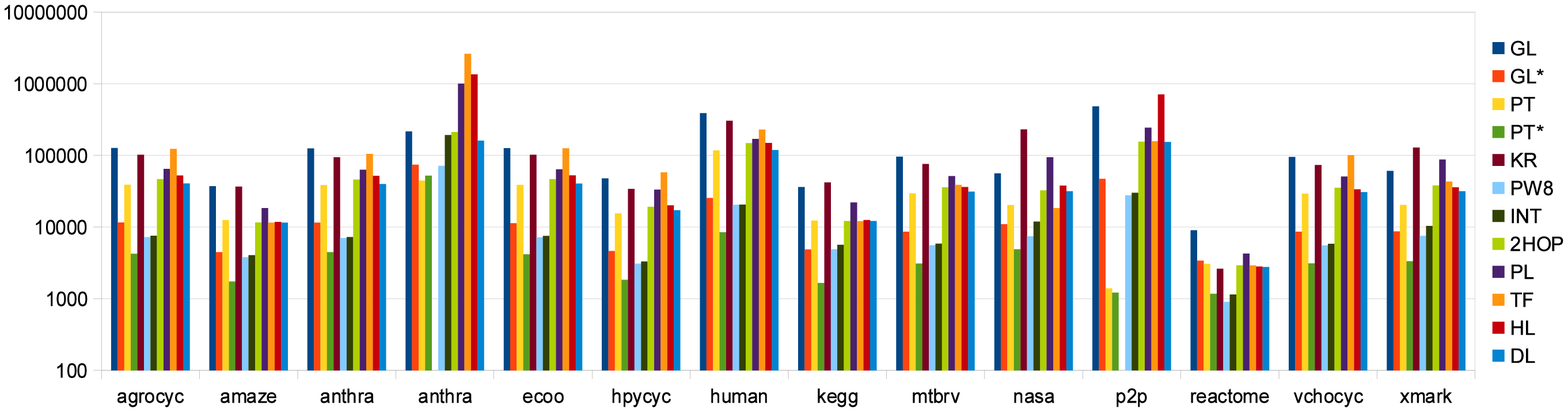,width=6.5in,height=1.8in}
	\vspace{-4.0ex}
	\caption {{\small Index Size on Small Real Graphs (in terms of the number of integers used in the indices)}}
	\label{fig:smallindex}
\vspace{1.0ex}
	\end{minipage}
%}

\resizebox{7.0in}{!} {
	\begin{minipage}[l]{6.5in}
	\centering
	\psfig{figure=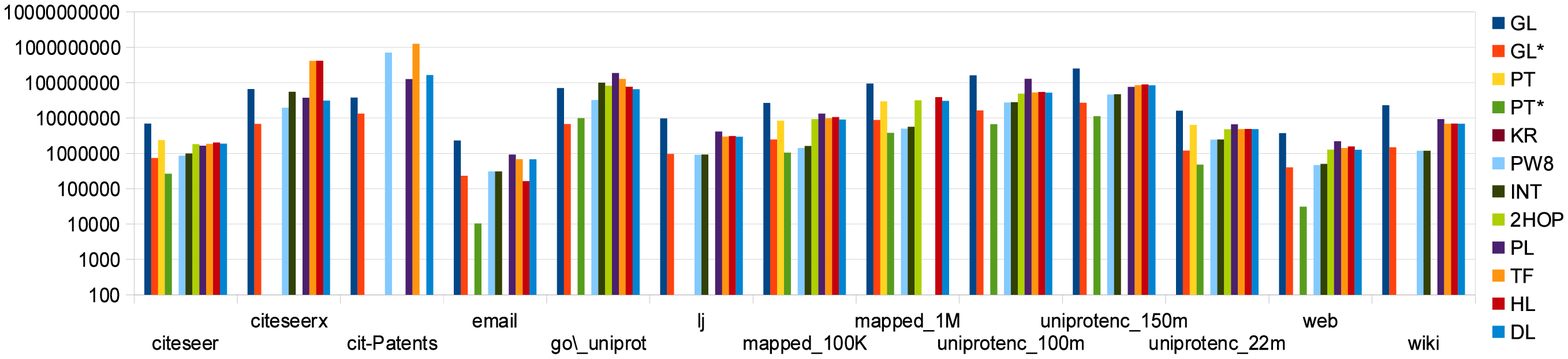,width=6.0in,height=1.6in}
	\vspace{-2.0ex}
	\caption {{\small Index Size on Large Real Graphs (in terms of the number of integers used in the indices)}}
	\label{fig:largeindex}
\vspace{-1.0ex}
	\end{minipage}
              \hspace{0.02cm}
}
\end{figure*}

\comment{
\begin{figure*}[!ht]
\resizebox{7.0in}{!} {
	\begin{minipage}[l]{3.5in}
	\centering
	\psfig{figure=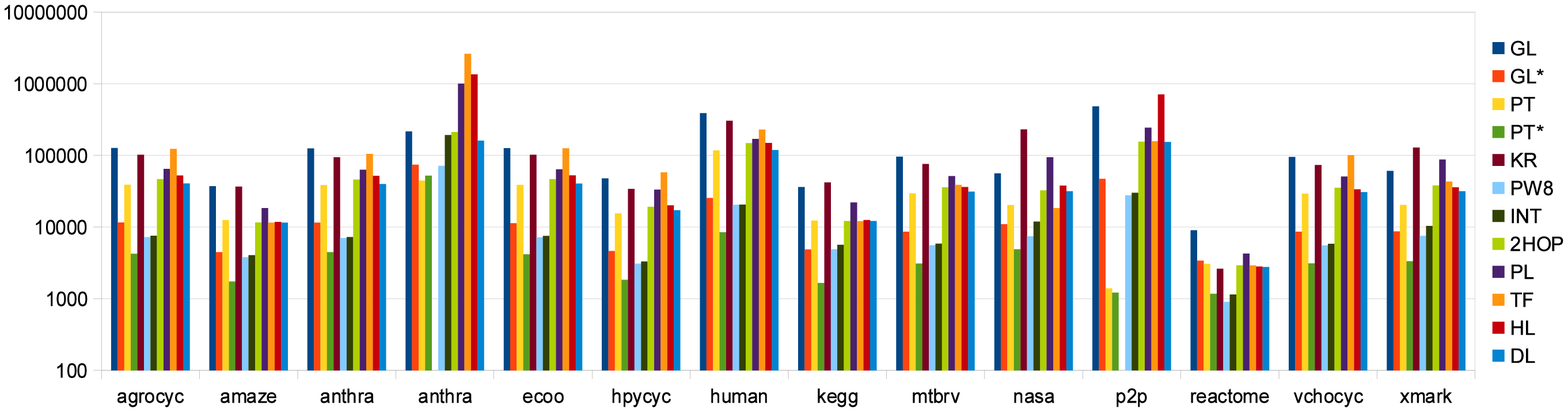,width=3.5in,height=4.0in}
	\vspace{-5.0ex}
	\caption {{\small Index Size on Small Real Graphs (in terms of the number of integers used in the indices)}}
	\label{fig:smallindex}
\vspace{-1.0ex}
	\end{minipage}
	\hspace{0.02cm}

	\begin{minipage}[l]{3.5in}
	\centering
	\psfig{figure=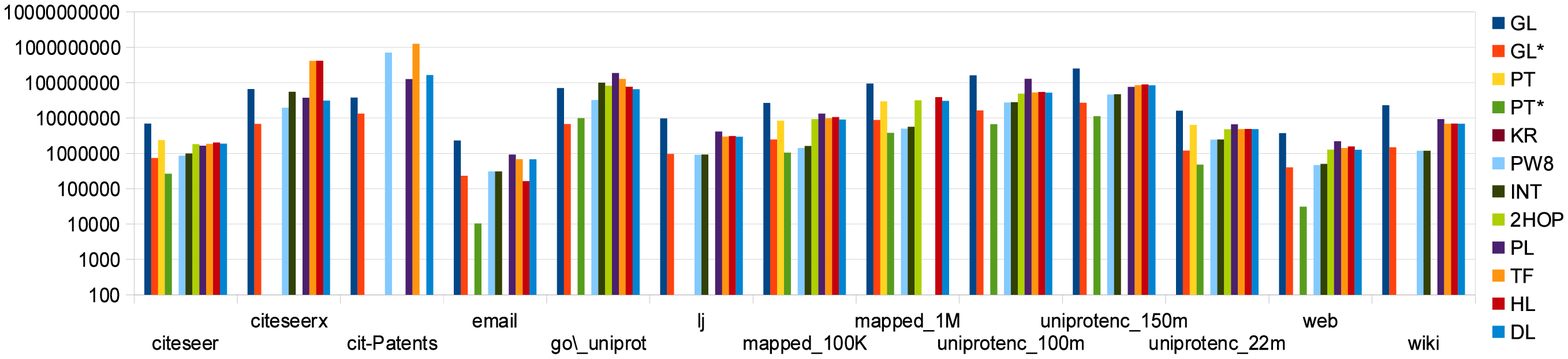,width=3.5in,height=4.0in}
	\vspace{-5.0ex}
	\caption {{\small Index Size on Large Real Graphs (in terms of the number of integers used in the indices)}}
	\label{fig:largeindex}
\vspace{-1.0ex}
	\end{minipage}
              \hspace{0.02cm}
}
\end{figure*}
}

\begin{table*}[!ht]
\centering
%\resizebox{7.0in}{!} {
{\small
\begin{tabular}{|l|r|r|r|r|r|r|r|r|r|r|r|r|}
\hline
\multicolumn{1}{|l|}{Dataset} &
\multicolumn{1}{l|}{GL} &
\multicolumn{1}{l|}{GL$^\ast$} &
\multicolumn{1}{l|}{PT} &
\multicolumn{1}{l|}{PT$^\ast$} &
\multicolumn{1}{l|}{KR} & \multicolumn{1}{l|}{PW8} & \multicolumn{1}{l|}{INT}&\multicolumn{1}{l|}{2HOP} & \multicolumn{1}{l|}{PL} & \multicolumn{1}{l|}{TF} & \multicolumn{1}{l|}{HL} & \multicolumn{1}{l|}{DL}
\\ \hline

citeseer	&	63.4 	&	42.4 	&	4.9 	&	26.9 	&	---	&	20.6 	&	12.3 	&	4.5 	&	82.3 	&	6.1 	&	7.7 	&	5.3 	\\ \hline
citeseerx	&	2012.3 	&	20230.9 	&	---	&	---	&	---	&	76.3 	&	8.8 	&	---	&	102.7 	&	37.2 	&	210.2 	&	7.7 	\\ \hline
cit-Patents	&	403.9 	&	711.1 	&	---	&	---	&	---	&	2538.9 	&	---	&	---	&	77.6 	&	56.7 	&	---	&	53.2 	\\ \hline
email	&	575.9 	&	30.2 	&	---	&	10.2 	&	---	&	13.8 	&	5.7 	&	---	&	88.4 	&	3.0 	&	2.2 	&	3.0 	\\ \hline
go\_uniprot	&	77.6 	&	80.4 	&	---	&	29.3 	&	---	&	41.9 	&	17.0 	&	16.0 	&	80.8 	&	1752.5 	&	6.2 	&	12.7 	\\ \hline
lj	&	11972.2 	&	2137.6 	&	---	&	---	&	---	&	10.4 	&	4.9 	&	---	&	28.9 	&	4.2 	&	4.7 	&	4.5 	\\ \hline
mapped\_100K	&	253.8 	&	92.3 	&	6.7 	&	25.1 	&	---	&	90.6 	&	6.0 	&	5.1 	&	51.9 	&	15.0 	&	5.1 	&	7.2 	\\ \hline
mapped\_1M	&	762.2 	&	99.0 	&	8.4 	&	26.4 	&	---	&	46.6 	&	6.3 	&	5.6 	&	---	&	---	&	6.1 	&	12.4 	\\ \hline
uniprotenc\_100m	&	82.7 	&	37.3 	&	---	&	31.9 	&	---	&	29.8 	&	19.8 	&	---	&	157.4 	&	9.3 	&	5.7 	&	9.0 	\\ \hline
uniprotenc\_150m	&	79.1 	&	45.6 	&	---	&	35.8 	&	---	&	31.1 	&	20.3 	&	---	&	159.2 	&	16.0 	&	6.5 	&	7.8 	\\ \hline
uniprotenc\_22m	&	53.0 	&	23.6 	&	6.2 	&	23.7 	&	---	&	23.1 	&	15.5 	&	47598.4 	&	81.8 	&	6.2 	&	4.4 	&	5.9 	\\ \hline
web	&	2369.6 	&	1041.9 	&	---	&	11.6 	&	---	&	13.1 	&	5.5 	&	3.0 	&	12.5 	&	4.2 	&	3.5 	&	3.9 	\\ \hline
wiki	&	100313.0 	&	25655.5 	&	---	&	---	&	---	&	8.6 	&	6.0 	&	---	&	179.2 	&	4.5 	&	3.6 	&	4.9 	\\ \hline

\end{tabular}
}
%}
\vspace{-2.0ex}
\caption{Query Time (ms) Based on Equal Query of Large Real Datasets}
\label{largerealdataEqualquery}
\vspace*{-1.0ex}
\end{table*}

\begin{table*}[!ht]
\centering
%\resizebox{7.0in}{!} {
{\small
\begin{tabular}{|l|r|r|r|r|r|r|r|r|r|r|r|r|}
\hline
\multicolumn{1}{|l|}{Dataset} &
\multicolumn{1}{l|}{GL} &
\multicolumn{1}{l|}{GL$^\ast$} &
\multicolumn{1}{l|}{PT} &
\multicolumn{1}{l|}{PT$^\ast$} &
\multicolumn{1}{l|}{KR} & \multicolumn{1}{l|}{PW8} & \multicolumn{1}{l|}{INT}&\multicolumn{1}{l|}{2HOP} & \multicolumn{1}{l|}{PL} & \multicolumn{1}{l|}{TF} & \multicolumn{1}{l|}{HL} & \multicolumn{1}{l|}{DL}
\\ \hline

citeseer	&	40.2 	&	21.4 	&	4.4 	&	22.6 	&	---	&	12.4 	&	9.6 	&	7.0 	&	137.6 	&	1.2 	&	4.7 	&	7.1 	\\ \hline
citeseerx	&	2585.6 	&	719.1 	&	---	&	---	&	---	&	39.8 	&	13.4 	&	---	&	163.1 	&	55.4 	&	23.7 	&	11.9 	\\ \hline
cit-Patents	&	501.5 	&	517.2 	&	---	&	---	&	---	&	1766.3 	&	---	&	---	&	201.9 	&	66.5 	&	---	&	48.1 	\\ \hline
email	&	754.4 	&	8.3 	&	---	&	11.3 	&	---	&	14.0 	&	10.1 	&	---	&	126.3 	&	5.8 	&	3.5 	&	5.0 	\\ \hline
go\_uniprot	&	47.6 	&	29.8 	&	---	&	26.2 	&	---	&	52.5 	&	20.8 	&	13.0 	&	120.2 	&	18.7 	&	12.0 	&	23.5 	\\ \hline
lj	&	829613.0 	&	448.9 	&	---	&	---	&	---	&	21.2 	&	11.7 	&	---	&	69.3 	&	8.2 	&	5.7 	&	7.6 	\\ \hline
mapped\_100K	&	52.4 	&	23.6 	&	5.9 	&	20.8 	&	---	&	4.9 	&	5.0 	&	6.5 	&	102.0 	&	7.2 	&	6.7 	&	9.4 	\\ \hline
mapped\_1M	&	55.0 	&	24.7 	&	8.7 	&	23.9 	&	---	&	5.6 	&	6.7 	&	7.1 	&	---	&	---	&	9.8 	&	9.9 	\\ \hline
uniprotenc\_100m	&	53.0 	&	33.6 	&	---	&	29.7 	&	---	&	28.3 	&	20.1 	&	---	&	212.7 	&	14.2 	&	7.5 	&	10.8 	\\ \hline
uniprotenc\_150m	&	56.6 	&	33.0 	&	---	&	31.9 	&	---	&	29.1 	&	23.1 	&	---	&	201.5 	&	16.0 	&	10.7 	&	11.3 	\\ \hline
uniprotenc\_22m	&	40.5 	&	25.6 	&	9.1 	&	24.8 	&	---	&	21.9 	&	15.2 	&	4.4 	&	116.4 	&	7.7 	&	5.9 	&	8.7 	\\ \hline
web	&	61295.3 	&	386.8 	&	---	&	18.5 	&	---	&	22.3 	&	9.2 	&	4.4 	&	53.9 	&	8.3 	&	4.7 	&	6.3 	\\ \hline
wiki	&	76336.7 	&	28.2 	&	---	&	---	&	---	&	6.2 	&	8.8 	&	---	&	270.7 	&	8.1 	&	5.9 	&	9.3 	\\ \hline

\end{tabular}
%}
}
\vspace{-2.0ex}
\caption{Query Time (ms) Based on Random Query of Large Real Datasets}
\label{largerealdataRandomquery}
\vspace*{-1.0ex}
\end{table*}

\begin{table*}[!ht]
\centering
%\resizebox{7.0in}{!} {
{\small
\begin{tabular}{|l|r|r|r|r|r|r|r|r|r|r|r|r|}
\hline
\multicolumn{1}{|l|}{Dataset} &
\multicolumn{1}{l|}{GL} &
\multicolumn{1}{l|}{GL$^\ast$} &
\multicolumn{1}{l|}{PT} &
\multicolumn{1}{l|}{PT$^\ast$} &
\multicolumn{1}{l|}{KR} & \multicolumn{1}{l|}{PW8} & \multicolumn{1}{l|}{INT}&\multicolumn{1}{l|}{2HOP} & \multicolumn{1}{l|}{PL} & \multicolumn{1}{l|}{TF} & \multicolumn{1}{l|}{HL} & \multicolumn{1}{l|}{DL}
\\ \hline

citeseer	&	2,011 	&	189 	&	18,025 	&	1,131 	&	---	&	487 	&	307 	&	14,054 	&	610 	&	677 	&	2,232 	&	528 	\\ \hline
citeseerx	&	17,564 	&	3,734 	&	---	&	---	&	---	&	17,006 	&	7,015 	&	---	&	12,670 	&	206,965 	&	182,068 	&	9,909 	\\ \hline
cit-Patents	&	15,669 	&	7,693 	&	---	&	---	&	---	&	935,457 	&	---	&	---	&	28,314 	&	79,412 	&	---	&	114,583 	\\ \hline
email	&	591 	&	54 	&	---	&	1,218 	&	---	&	213 	&	136 	&	---	&	180	&	124 	&	211 	&	131 	\\ \hline
go\_uniprot	&	32,358 	&	3,163 	&	---	&	5,038,350 	&	---	&	34,373 	&	20,664 	&	252,540 	&	27,762 	&	57,761 	&	279,132 	&	16,706 	\\ \hline
lj	&	2,603 	&	213 	&	---	&	---	&	---	&	670 	&	786 	&	---	&	856 	&	551 	&	1,195 	&	313 	\\ \hline
mapped\_100K	&	6,220 	&	692 	&	26,667 	&	5,105 	&	---	&	448 	&	419 	&	9,760 	&	2,424 	&	1,951 	&	10,141 	&	1,902 	\\ \hline
mapped\_1M	&	28,303 	&	2,791 	&	103,265 	&	23,119 	&	---	&	2,399 	&	3,777 	&	52,190 	&	---	&	---	&	45,490 	&	6,894 	\\ \hline
uniprotenc\_100m	&	66,285 	&	5,090 	&	---	&	8,168,170 	&	---	&	16,330 	&	11,624 	&	1,028,970 	&	13,882 	&	49,448 	&	67,270 	&	13,854 	\\ \hline
uniprotenc\_150m	&	101,556 	&	8,695 	&	---	&	18,840,100 	&	---	&	27,202 	&	18,863 	&	---	&	16,332 	&	34,245 	&	119,570 	&	21,015 	\\ \hline
uniprotenc\_22m	&	5,034 	&	300 	&	9,801,660 	&	41,204 	&	---	&	1,408 	&	1,064 	&	102,679 	&	1184 	&	1,970 	&	5,209 	&	1,004 	\\ \hline
web	&	953 	&	72 	&	---	&	46,247 	&	---	&	612 	&	1,812 	&	32,017,700 	&	474 	&	390 	&	1,352 	&	627 	\\ \hline
wiki	&	7,063 	&	319 	&	---	&	---	&	---	&	432 	&	698 	&	---	&	1,763 	&	902 	&	5,398 	&	1,527 	\\ \hline

\end{tabular}
}
%}
\vspace{-2.0ex}
\caption{Construction Time (ms) of Large Real Datasets}
\label{largerealdataConstruction}
\vspace*{-1.0ex}
\end{table*}

\comment{
\begin{figure*}[!ht]
\\resizebox{7.5in}{!} {
	\begin{minipage}[c]{4.5in}
	\centering
	\psfig{figure=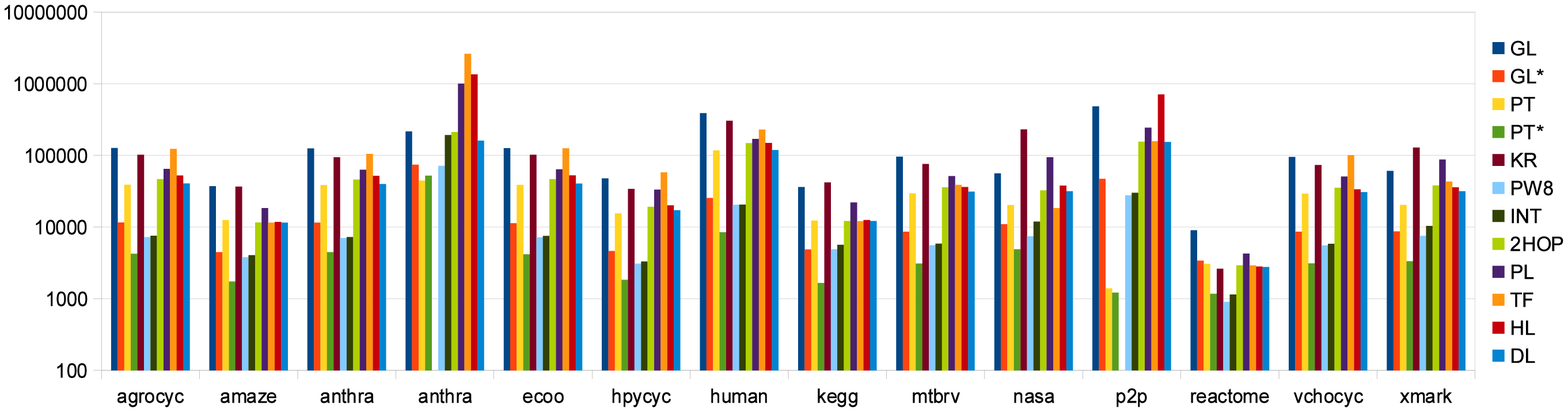,width=4.5in,height=2.0in}
	\vspace{-3.0ex}
	\caption {{\small Construction Time (ms) on Real Small Graphs}}
	\label{fig:constimesmall}
\vspace{-3.0ex}
	\end{minipage}
	\hspace{0.02cm}

	\begin{minipage}[l]{3.0in}
	\centering
	\psfig{figure=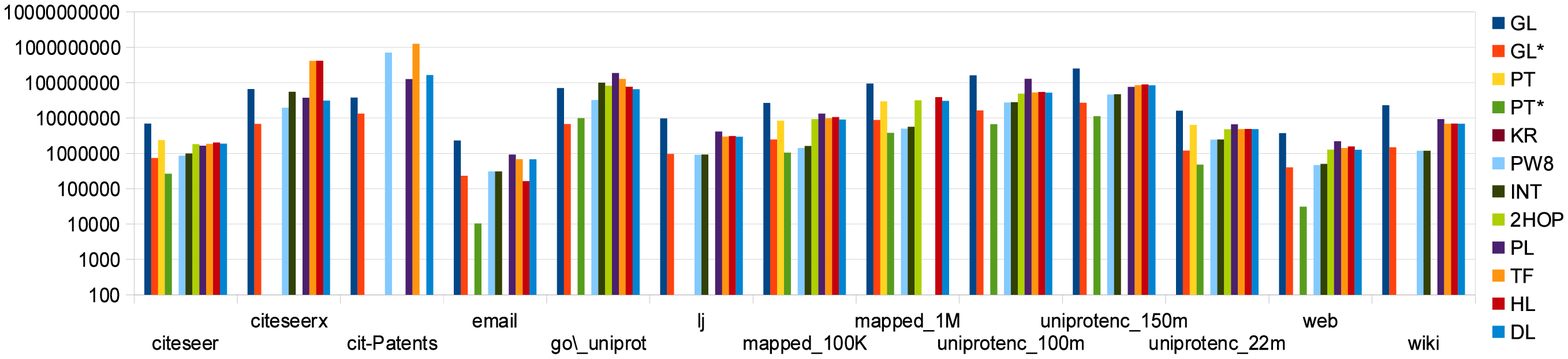,width=3.0in,height=2.0in}
	\vspace{-3.0ex}
	\caption {{\small Construction Time (ms) on Real Large Graphs}}
	\label{fig:constimelarge}
\vspace{-3.0ex}
	\end{minipage}
            \hspace{0.02cm}
}
\end{figure*}
}

We make the following important observations on the query time:
1) On small graphs, PATH-TREE outperforms other methods, though K-REACH is fairly close (as it is quite similar to the transitive closure materialization). Interestingly, the reachability oracle methods turn out to be quite comparable. In particular, the Distribution-Labeling (DL) is consistently about $2$ times slower than PATH-TREE, and even faster than the other transitive closure compression approaches, INTERVAL and PWAH-8, on equal query load. \\
2) Compared to the existing set-cover based labeling approach 2HOP, Hierarchical-Labeling (HL) is quite comparable (slightly slower), but the query time of Distribution-Labeling (DL) is only $2/3$ of that of 2HOP. \\
3) The reachability oracle approaches are slightly slower on the random query load than on the equal query load.
This is because to determine vertex $u$ cannot reach vertex $v$, the query processing has to completely scan $L_{out}(u)$ and $L_{in}(v)$.
4) Due to additional distance comparison cost, the Pruned Landmark (PL) is fairly slow; its query performance is close to the GRAIL.
5) Both Hierarchical Labeling (HL) and Distribution Labeling (DL) are faster than the TF-labeling (TF); though they are quite comparable when the workload is random. We analyzed the source code of TF and found it utilizes some additional optimization technique (such as using topological order) to quickly reject non-reachable pairs. This can be a potential reason for its handling for random query load which mainly consists of non-reachable pairs. 

Table~\ref{smallrealdataConstruction} shows the construction time of different reachability indices on small graphs.
We observe K-REACH and 2HOP are the slowest. This is understandable as K-REACH needs to perform vertex-cover discovery and materialize the transitive closure for the vertex-cover; and 2HOP needs to perform the expensive greedy set-cover and completely materialize the transitive closure. INTERVAL and PAWH-8 turn out be the fastest and even faster than the online search GRAIL approach as the later still needs to perform random DFS a few times (in this study, we choose the number to be $5$ as being used in~\cite{yildirim:grail}). Both Hierarchical-Labeling (HL) and Distribution-Labeling (DL) are much more efficient in labeling: The Hierarchical-Labeling is on average $5$ times faster than 2HOP whereas the Distribution-Labeling is consistently $20$ times faster (and in some case more than two order of magnitude faster) than 2HOP. In fact, it has even faster construction time than GRAIL and quite comparable to the INTERVAL and PWAH-8.
The TF-labeling (TF) and the Pruned Landmark (PL)  are typically faster than Hierarchical-Labeling (HL) but slower than  Distribution-Labeling (DL). This is understandable as TF is simpler than Hierarchical-Labeling (HL) and PL needs additional computation cost with respect to  Distribution-Labeling (DL). 

Figure~\ref{fig:smallindex} shows the index size of different reachability index methods along with some of their  SCARAB counterparts on small graphs.
Here, PWAH-8 and INTERVAL outperform the others on index size. It is interestingly to observe that the labeling size of Hierarchical-Labeling (HL) is quite comparable to 2HOP (and this is also consistent with the query time).
More importantly and rather surprisingly, the labeling size of Distribution-Labeling (DL) is consistently smaller than that of 2HOP, the set-cover based optimization labeling targeting for minimizing the labeling size.
This, we believe, can be attributed to the effectiveness of the total order based hierarchy and the non-redundant labeling process.
Also, both Distribution Labeling (DL) and Hierarchical Labeling (HL) produce smaller than labeling size than TF-labeling (TF). This demonstrates the advantage of using reachability backbone and also directly contributes to the faster query performance for HL (and DL). 

\noindent{\bf Large Graphs:}
Large graphs provide the real challenge for the reachability computation.
We observe that only three methods, GRAIL, PWAH-8, and Distribution-Labeling are able to handle all these graphs (GRAIL$^\ast$ is the SCARAB variant for speeding up query performance). Hierarchical-Labeling, Pruned Landmark, INTERVAL, and TF-Labeling can work on $12$, and PATH-TREE$^\ast$ can work on $9$, out of $13$ large graphs.
K-REACH, PATH-TREE and 2-HOP fails on most of the large graphs.  For K-REACH and PATH-TREE, their labeling size are too large to be materialized in the main memory; for 2-HOP, its running time  for these graphs often exceeds the 24-hour time limit. 

%K-REACH can only perform one graph, where PATH-TREE and 2HOP fail on $5$ and $4$ large graphs, respectively.

Tables~\ref{largerealdataEqualquery} and \ref{largerealdataRandomquery}
report the query time using the {\em equal} and {\em random} query load, respectively.
We make the following observations: 1)  On large graphs, the transitive closure compression approaches, even on the graphs they can work, become significant slower. This is expected as the compressed transitive closure $TC(v)$ becomes larger, its search (linear or binary) becomes more expensive. Now, the advantage of the reachability oracle becomes clear as they become the fastest in terms of query time (even faster than PATH-TREE and INTERVAL, and consistently more than $5$ times faster than PWAH-8). 2) Compared with the original 2HOP labeling, both Hierarchical-Labeling and Distribution-Labeling have comparable query performance on the graphs which they all can run. 3) The latest TF-labeling (TF) is  slower than both  Hierarchical-Labeling (HL) and Distribution-Labeling (DL) on most of the large graphs and for both equal and random query work load. 

Tables~\ref{largerealdataConstruction} shows the construction time on large graphs for all methods. We observe that PAWH-8 and INTERVAL are very fast though as the graph becomes larger, they become slower or cannot finish. Distribution-Labeling turns out to be quite comparable (fastest on several graphs). Hierarchical-Labeling can work on $8$ out of $9$ graphs and it shows signifiant improvement on $2$ out of $5$ graphs which 2HOP can also process. Distribution-Labeling is on average of one order of magnitude performance faster than 2HOP on these five graphs.
Also, on average, the construction time of TF-labeling (TF)  is comparable to that of Hierarchical-Labeling (HL), and is slower than that of  Distribution-Labeling (DL).  

Figure~\ref{fig:largeindex} shows the index size of different approaches. The results are quite consistent with the results on the small graphs on those graphs they can work. For most cases, PWAH-8 And INTERVAL have the smallest index size. 2HOP, Hierarchical-Labeling and Distribution-Labeling also perform well (better than GRAIL and K-Reach). The labeling sizes of 2HOP, Hierarchical-Labeling and Distribution-Labeling are quite comparable; Distribution-Labeling has smaller labeling size than Hierarchical Labeling and very close to (or better than) 2HOP on the graphs it can run.
Finally, on average, the index size of TF-labeling is quite close to that of Hierarchical-Labeling (HL), but slightly higher than Distribution-Labeling (DL).

%We note that we also evaluated these methods on scale-free directed graphs with  the number of vertices ranging from $1M$ vertices to $10M$ vertices, and the average edge density is fixed to be $2$. The results are largely consistent with the results on the large real graphs. Due to the space limitation, we do not report the detailed results on these graphs.

\comment{
Table~\ref{syndata} shows the number of vertices and edges in $G^\ast$ with respect to $\epsilon$ on these large graphs.
Clearly, as $\epsilon$ increases, the size of backbone also reduces. However, the density of the backbone will also increase accordingly.
This is expected as the local search range for each backbone vertex increases.
But we note that the increase rate is rather small.
Interestingly, we note that even when $\epsilon=2$, the vertex reduction rate is around $4$ to $5$ times.
Also, this seems to be lower than the real graphs.
This suggests that there are still some important properties of the real world graphs, which are not well captured by the existing graph model, such as the scale-free (power-law degree distribution) model.
Finally, we note that the INTERVAL approach can handle the graph up to $3,000,000$ vertices.
Using the {\bf SCARAB} framework, it can handle all these graphs, and its query performance can be more than one order of magnitude faster than GRAIL.
Duo to the space limitation, we do not report the query time, index size, and construction time on these graphs.

}

\vspace*{-2.0ex}
\section{Conclusion}
\label{conc}

In this paper, by introducing two simple, elegant, and effective labeling approaches, {\em Hierarchical Labeling} and {\em Distribution Labeling}, we are able to resolve an important open question in reachability computation: the reachability oracle can be a powerful tool (or even the most useful one) to handle real large graphs. Our experimental results demonstrate that they can perform on graphs with millions of vertices/edges (scalable), are quickest in answering reachability queries on large graphs (fast), and have comparable or better labeling size as the set-cover based optimization approaches (compact). 
In the future, we will investigate the labeling on dynamic graphs and how to apply them on more general reachability computation, such as $k$-reach problem.

%\setlength{\bibsep}{0.0pt} % need to \usepackage{natbib} for this
%\begin{scriptsize}
%\begin{spacing}{0.7}
\begin{small}
\bibliographystyle{plain}
\bibliography{../bib/distance,../bib/reachabilitytods,../bib/reachpaper,../bib/Proposal,../bib/3hop,../bib/paper,../bib/cikm,../bib/reachability,../bib/ComplexNetwork2}
\end{small}
%\end{spacing}
%\end{scriptsize}
%\newpage
%\input{appendix}

\end{document}